
\input harvmac
\noblackbox
\def\Im{{\rm Im}}
\def\Re{{\rm Re}}
\def\cof{{\rm cof}}

\def\mod{{\rm mod}}
\def\evec{{\bf e}}
\def\difftau{(\tau-\bar\tau)}
\def\diag{{\rm diag}}
\def\ncal{{\cal N}}

\lref\DFhananykol{
A.~Hanany and B.~Kol,
``On orientifolds, discrete torsion, branes and M theory,''
JHEP {\bf 0006}, 013 (2000)
[arXiv:hep-th/0003025].
}
\lref\DFwitten{
E.~Witten,
``Toroidal Compactification without Vector Structure,''
JHEP {\bf 9802}, 006 (1998)
[arXiv:hep-th/9712028]. 
}
\lref\bergsh{E. Bergshoeff, M. de Roo and E. Eyras, ``Gauged
Supergravity from Dimensional Reduction,'' Phys. Lett.
{\bf B413}, 70 (1997) [arXiv:hep-th/9707130].}
\lref\Myers{R. Myers, ``Dielectric Branes,'' JHEP {\bf 9912}, 022
(1999) [arXiv:hep-th/9910053].}
\lref\oldflux{S. Gukov, C. Vafa and E. Witten,
``CFTs from Calabi-Yau Fourfolds,'' Nucl. Phys. {\bf B584}, 69
(2000) [arXiv:hep-th/9906070]\semi
B. Greene, K. Schalm and G. Shiu, ``Warped Compactifications in
M and F-theory,'' Nucl. Phys. {\bf B584}, 480 (2000)
[arXiv:hep-th/0004103].}

\lref\DRS{K. Dasgupta, G. Rajesh and S. Sethi, ``M-theory, Orientifolds
and G-flux,'' JHEP {\bf 9908}, 023 (1999) [arXiv:hep-th/9908088].}

\lref\SVW{S. Sethi, C. Vafa and E. Witten, ``Constraints on Low
Dimensional String Compactifications,'' Nucl. Phys. {\bf B480},
213 (1996) [arXiv:hep-th/9606122].}

\lref\kalmy{N. Kaloper and R. Myers, ``The O(d,d) Story
of Massive Supergravity,'' JHEP {\bf 9905}, 010 (1999)
[arXiv:hep-th/9901045].}

\lref\asym{

J. Harvey, G. Moore and C. Vafa, ``Quasicrystalline Compactification,''
Nucl. Phys. {\bf B304}, 269 (1988)\semi
M. Dine and E. Silverstein, ``New M Theory Backgrounds with Frozen
Moduli,'' [arXiv:hep-th/9712166]\semi
A. Dabholkar and J. Harvey, ``String Islands,'' JHEP {\bf 9902},
006 (1999) [arXiv:hep-th/9809122].}
\lref\Beckers{K. Becker and M. Becker, ``M-theory on Eight Manifolds,''
Nucl. Phys. {\bf B477}, 155 (1996) [arXiv:hep-th/9605053].}
\lref\Beckerst{
K. Becker and M. Becker, ``Supersymmetry Breaking, M-theory
and Fluxes,'' JHEP {\bf 0107}, 038 (2001) [arXiv:hep-th/0107044].}

\lref\KPV{S. Kachru, J. Pearson and H. Verlinde, ``Brane/Flux Annihilation
and the String Dual of a Non-Supersymmetric Field Theory,'' [arXiv:hep-th/0112197].}

\lref\louis{M. Haack and J. Louis, ``M-theory Compactified on a
Calabi-Yau Fourfold with Background Flux,'' Phys. Lett. {\bf B507},
296 (2001) [arXiv:hep-th/0103068].}

\lref\louishet{J. Louis and A. Micu, ``Heterotic String Theory
with Background Fluxes,'' [arXiv:hep-th/0110187].}

\lref\curio{G. Curio, A. Klemm, D. Lust and S. Theisen, ``On the
Vacuum Structure of Type II String Compactifications on Calabi-Yau
Spaces with H Fluxes,'' Nucl. Phys. {\bf B609}, 3 (2001) [arXiv:
hep-th/0012213].}

\lref\dasflux{K. Dasgupta, K. Oh, J. Park and R. Tatar, ``Geometric
Transition versus Cascading Solution,'' [arXiv:hep-th/0110050].}

\lref\EvaFix{E. Silverstein, ``(A)dS Backgrounds from Asymmetric
Orientifolds,'' [arXiv:hep-th/0106209].}

\lref\JoeOne{
J.~Polchinski,
``String Theory. Vol. 1: An Introduction To The Bosonic String,''
{\it  Cambridge, UK: Univ. Pr. (1998) 402 p}.
}
\lref\JoeTwo{
J.~Polchinski,
``String Theory. Vol. 2: Superstring Theory And Beyond,''
{\it  Cambridge, UK: Univ. Pr. (1998) 531 p}.
}

\lref\joefrey{A. Frey and J. Polchinski, ``${\cal N}=3$ Warped
Compactifications,'' preprint to appear.}

\lref\GKP{
S.~B.~Giddings, S.~Kachru and J.~Polchinski,
``Hierarchies from fluxes in string compactifications,''
[arXiv:hep-th/0105097].
}
\lref\GPOne{
M.~Grana and J.~Polchinski,
``Gauge/gravity duals with holomorphic dilaton,''
[arXiv:hep-th/0106014].
}
\lref\GPTwo{
M.~Grana and J.~Polchinski,
``Supersymmetric three-form flux perturbations on AdS(5),''
Phys.\ Rev.\ D {\bf 63}, 026001 (2001)
[arXiv:hep-th/0009211].
}

\lref\Curiodual{G. Curio, A. Klemm, B. Kors and D. Lust, ``Fluxes in
Heterotic and Type II String Compactifications,'' [arXiv:hep-th/0106155].} 

\lref\andyjoe{J. Polchinski and A. Strominger, ``New Vacua for Type II String
Theory,'' Phys. Lett. {\bf B388}, 736 (1996) [arXiv:hep-th/9510227].}

\lref\mich{J. Michelson, ``Compactification of Type IIB Strings to Four-Dimensions
with Nontrivial Classical Potential,'' Nucl. Phys. {\bf B495}, 127 (1997)
[arXiv:hep-th/9610151].} 

\lref\ferrara{L. Andrianopoli, R. D'Auria and S. Ferrara, ``Consistent reduction
of ${\cal N}=2$ $\to$ ${\cal N}=1$ four-dimensional supergravity coupled
to matter,'' [arXiv:hep-th/0112192].} 

\lref\dallgata{G. Dall'Agata, ``Type IIB supergravity compactified on a Calabi-Yau
manifold with H-fluxes,'' [arXiv:hep-th/0107264].}

\lref\Moore{
G.~W.~Moore,
``Arithmetic and attractors,''
[arXiv:hep-th/9807087].
}

\lref\Mayr{P. Mayr, ``On Supersymmetry Breaking in String Theory
and its Realization in Brane Worlds,'' Nucl. Phys. {\bf B593}, 99
(2001) [arXiv:hep-th/0003198]\semi
P. Mayr, ``Stringy Brane Worlds and Exponential Hierarchies,''
JHEP {\bf 0011}, 013 (2000), [arXiv:hep-th/0006204].}

\lref\PS{J. Polchinski and M. Strassler, ``The String Dual of a
Confining Four-Dimensional Gauge Theory,'' [arXiv:hep-th/0003136].}

\lref\RS{L. Randall and R. Sundrum, ``A Large Mass Hierarchy
from a Small Extra Dimension,'' Phys. Rev. Lett. {\bf 83}, 3370
(1999) [arXiv:hep-th/9905221].}

\lref\Herman{H. Verlinde, ``Holography and Compactification,''
Nucl. Phys. {\bf B580}, 264 (2000) [arXiv:hep-th/9906182]\semi
C. Chan, P. Paul and H. Verlinde, ``A Note on Warped String
Compactification,'' Nucl. Phys. {\bf B581}, 156 (2000)
[arXiv:hep-th/0003236].}

\lref\Hitchin{N. Hitchin, ``The geometry of three-forms in six and seven
dimensions,'' [arXiv:DG/0010054].}

\lref\Georgi{H.~Georgi, ``Lie Algebras in Particle Physics,''{\it Frontiers
in Physics}, ISBN 0-8053-3153-0.} 

\lref\Schlich{
M.~Schlichenmaier
``An Introduction to Riemann Surfaces, Algebraic Curves and Moduli Spaces,''
{\it  Berlin: Springer-Verlag (1989) 148 p}, (Lecture Notes in Physics, 322).
}

\Title{\vbox{\baselineskip12pt
\hbox{hep-th/0201028}
\hbox{SU-ITP-01/49}
\hbox{SLAC-PUB-9066}
\hbox{TIFR-TH/01-51}
}}
{\vbox{\centerline{Moduli Stabilization from Fluxes}
\vskip2pt\centerline{in a Simple IIB Orientifold}}}

\centerline{Shamit Kachru $^{a}$\footnote{$^1$}{skachru@stanford.edu},
Michael Schulz $^{a}$\footnote{$^2$}{mschulz@stanford.edu} and
Sandip P. Trivedi $^{b}$\footnote{$^3$}{sandip@tifr.res.in}}
\medskip\centerline{$^a$\it Department of Physics and SLAC, Stanford University}
\centerline{\it Stanford, CA 94305/94309, USA}
\medskip
\centerline{$^b$\it Tata Institute of Fundamental Research}
\centerline{\it Homi Bhabha Road, Mumbai 400 005, INDIA}

\vskip .15in
We study novel type IIB compactifications on the $T^6/Z_2$
orientifold.  This geometry arises in the T-dual description of Type I
theory on $T^6$, and one normally introduces 16 space-filling
D3-branes to cancel the RR tadpoles.  Here, we cancel the RR tadpoles
either partially or fully by turning on three-form flux in the compact
geometry.  The resulting (super)potential for moduli is calculable.
We demonstrate that one can find many examples of ${\cal N}=1$
supersymmetric vacua with greatly reduced numbers of moduli in this
system.  A few examples with ${\cal N}>1$ supersymmetry or complete
supersymmetry breaking are also discussed.

\Date{January 2002}

\newsec{Introduction}

The study of Calabi-Yau orientifold compactifications of type II
string theory (or F-theory compactifications on Calabi-Yau fourfolds),
with nontrivial background RR and NS fluxes through compact cycles of
the Calabi-Yau manifold, is of interest for several reasons.

Conventional compactifications give rise to models which typically
have many moduli. Understanding how these flat directions are lifted
is important, both from the point of view of phenomenology and of
cosmology.  One expects the moduli to be lifted once supersymmetry is
broken, but studying this in a calculable way in conventional
compactifications has proved challenging so far.  In contrast,
compactifications with background RR and NS fluxes turned on give rise
to a nontrivial low energy potential which freezes many of the
Calabi-Yau moduli. Moreover, the potential is often calculable and as
a result one can hope to study the stabilization of many moduli in a
controlled manner in this setting.  Flux-induced potentials for moduli
have been discussed before in
e.g. \refs{\andyjoe,\mich,\oldflux,\DRS,\GKP,\EvaFix,\curio} (while a
complementary ``stringy'' means of freezing moduli, by considering
asymmetric orbifolds, has been discussed in, for example, \asym).

Compactifications with fluxes have also been proposed as a natural
setting for warped solutions to the hierarchy problem \Herman, along
the lines of the proposal of Randall and Sundrum \RS. The combination
of fluxes and space filling D-branes which often need to be introduced
for tadpole cancellation in these models leads to a nontrivial warped
metric, with the scale of 4d Minkowski space varying over the compact
dimensions.  Examples of such models, with almost all moduli
stabilized and exponentially large warping giving rise to a hierarchy,
appeared in \GKP. (See also \Mayr).

Finally, compactifications with fluxes also have interesting (and
relatively unexplored) dual descriptions, via mirror symmetry and
heterotic/type II duality.  Some examples of these dualities have been
discussed in \refs{\DRS,\Curiodual}.

In this paper, we explore in detail the simplest such compactification
which admits supersymmetric vacua with nontrivial NS and RR fluxes:
the compactification of type IIB string theory on the $T^6/Z_2$
orientifold.  The most familiar avatar of this model includes 16
D3-branes which cancel the RR charge of the 64 O3-planes at the $2^6$
fixed points of the $Z_2$ action.  However, one is free to replace
some (or all) of the D3-branes with appropriate integral RR and NS
3-form fluxes $F_{(3)}$ and $H_{(3)}$.  Given such a choice of
integral fluxes, one can compute the low-energy superpotential
governing the light fields.  In a generic Calabi-Yau orientifold in
IIB string theory, the periods which are required to determine $W$
would only be computable as approximate expansions about various
extreme points in moduli space, making any global and tractable
expression for $W$ difficult to obtain.  A nice feature of the
$T^6/Z_2$ case is that $W$ is easily computable.

With the superpotential in control we can ask if there are ${\cal
N}=1$ supersymmetry preserving minima. It turns out that for generic
choices of the fluxes supersymmetry is broken. By suitably choosing
the fluxes, however, we find several examples which give rise to
stable, ${\cal N}=1$ supersymmetric ground states.  In these minima,
typically, the dilaton-axion, all complex structure moduli, and some
of the K\"ahler moduli are stabilized. In addition, since some or all
of the O3-plane charge is cancelled by the flux, fewer D3-branes are
present, and the number of moduli coming from the open string sector
is also reduced. The conventional IIB compactification on this
$T^6/Z_2$ orientifold has $67$ (complex) moduli\foot{The model has
$16$ D3-branes each of which give rise to $3$ moduli, in addition
there are 19 moduli coming from the closed string sector.}.  Once
fluxes are turned on, it is easy to find examples with far fewer
moduli ($\sim 3$ in the models we discuss here, and fewer in the class
of models described in \GKP).

The organization of this paper is as follows.  In \S2, we review basic
facts about vacua with flux and about the moduli of the $T^6/Z_2$
orientifold, and parametrize the possible choices of flux.  In \S3, we
discuss the constraints that must be imposed to find a supersymmetric
vacuum, following \refs{\Beckers,\GPOne}, and write down a formula for
the superpotential as a function of the $T^6$ moduli.  In \S4, we
exhibit many examples which lead to ${\cal N}=1$ supersymmetric
solutions. We also analyze some cases which turn out to have ${\ncal=
3}$ supersymmetry and make some comments about finding the most
general supersymmetric solution.  In \S5\ we discuss the conditions
under which two apparently distinct solutions are nevertheless
equivalent (using the reparametrization symmetries of the torus and
U-duality).  In \S6\ we describe how, starting from a supersymmetric
solution, additional physically distinct ones can be found using
rescalings and $GL(2,{\bf Z}) \times GL(6,{\bf Z})$ transformations.
In \S7, we derive the conditions which must be imposed on the
$G_{(3)}$ flux to find ${\cal N}=2$ supersymmetric solutions, and
consider one illustrative example. \S8 contains some examples of
nonsupersymmetric solutions.  In \S9\ we discuss the dynamics on the
D3 branes which one should insert into many of our vacua, to saturate
the D3 tadpole.  We close with a brief description of directions for
future research in \S10, and some important details are relegated to
appendices A-D.

While this work was in progress, we learned of a related work
exploring novel 4d ${\cal N}=3$ supersymmetric vacua which can be
found from special flux configurations on $T^6/Z_2$ \joefrey.  We are
grateful to the authors of \joefrey\ for providing us with an early
version of their paper, and for helpful comments.

\newsec{Preliminaries}

\subsec{D3-brane charge from 3-form flux}

The type IIB supergravity action in Einstein frame is \JoeTwo
\eqn\IIBsugra{\eqalign{S_{\rm IIB}
&= {1\over2{\kappa_{10}}^2} \int d^{10}x\sqrt{-g}\biggl(
R-{\partial_M\phi\partial^M\phi\over2(\Im\phi)^2}
-{G_{(3)}\cdot\bar G_{(3)}\over 2\cdot3!\Im\phi}
-{\tilde F_{(5)}{}^2\over4\cdot 5!}\biggr)\cr
&+{1\over2{\kappa_{10}}^2}\int
{C_{(4)}\wedge G_{(3)}\wedge\bar G_{(3)}\over 4 i \Im\phi}
+ S_{\rm local}.}}
Here,
\eqn\fielddefs{\phi= C_{(0)}+i/g_s,\qquad
G_{(3)}= F_{(3)}-\phi H_{(3)},}
and
\eqn\Ffive{
\tilde F_{(5)} = F_{(5)}-\half C_{(2)}\wedge H_{(3)}
+\half F_{(3)}\wedge B_{(2)}, \quad
{\rm with}\ *\tilde F_{(5)} = \tilde F_{(5)},}
where $F_{(3)}=dC_{(2)}$ and $H_{(3)}=dB_{(2)}$.  If one compactifies
on a six dimensional compact manifold, ${\cal M}_6$, and includes the
possibility of space-filling D3-branes and O3-planes, then the
equation of motion/Bianchi identity for the 5-form field strength is
\eqn\bianchi{d\tilde F_{(5)} = d*\tilde F_{(5)}
= H_{(3)}\wedge F_{(3)} + 2{\kappa_{10}}^2\mu_3
\rho_3^{\rm local}.}
Here $\mu_3$ is the charge density of a D3-brane and $\rho_3^{\rm
local}$ is the number density of local sources of D3-brane charge on
the compact manifold.  We can integrate this equation over ${\cal
M}_6$ to give the condition
\eqn\intbianchi{{1\over2{\kappa_{10}}^2\mu_3}
\int_{{\cal M}_6}H_{(3)}\wedge F_{(3)} + Q_3^{\rm local}=0.}
In condition \intbianchi, $Q_3^{\rm local}$ is the sum of contibutions
$+1$ for each D3-brane and $-1/4$ for each normal O3-plane.  As
discussed in \DFhananykol\ and \DFwitten, there are actually three
other types of O3-plane, each characterized by the presence of
discrete RR and/or NS flux at the orientifold plane.  These exotic
O3-planes each contribute $+1/4$ to $Q_3^{\rm local}$.

We will be interested in the case that ${\cal M}_6$ is the $T^6/Z_2$
orientifold.  There are $2^6$ O3-planes in this compactification, with
a total contribution of $-16+\half N_{{\rm O3}'}$ units of D3-brane
charge to $Q_3^{\rm local}$, where $N_{{\rm O3}'}$ is the number of
exotic O3-planes.  Therefore, \intbianchi\ takes the form
\eqna\twostat
$$ \ha N_{\rm flux} + N_{\rm D3} + \ha N_{{\rm O3}'}=16.\eqno\twostat a$$

Here
$$N_{\rm flux} = {1\over (2 \pi)^4 (\alpha')^2} \int_{T^6} H_{(3)}
\wedge F_{(3)}.\eqno\twostat b$$
The factor of $\half$ multiplying $N_{\rm flux}$ compensates for the
fact that the integration is over $T^6$ rather than $T^6/Z_2$.  We
have also replaced the prefactor, $1/(2{\kappa_{10}}^2\mu_3)$, with
its explicit value in terms of $\alpha'$.  It is clear from
\twostat{}\ that appropriately chosen three-form fluxes can carry
D3-brane charge.  The fluxes obey a quantization condition
\eqn\quantization{{1\over(2\pi)^2\alpha'}\int_\gamma F_{(3)}=
m_\gamma\in{\bf Z},\qquad {1\over(2\pi)^2\alpha'}\int_\gamma 
H_{(3)}=n_\gamma\in{\bf Z},}
where $\gamma$ is an arbitrary class in $H_3(T^6,{\bf Z})$.  There is
a subtlety in arguing that these are the correct quantization
conditions for $T^6/Z_2$ \foot{We are indebted to A. Frey and
J. Polchinski for pointing out this subtlety.}\joefrey.  This is
because there are additional three cycles in $T^6/Z_2$, which are not
present in the covering space $T^6$.  If some of the integers
$m_\gamma$ ($n_\gamma$) are odd, additional discrete RR (NS) flux
needs to be turned on at appropriately chosen orientifold planes to
meet the quantization condition on these additional cycles.  (See
Appendix A for more discussion of this condition).  In practice it is
quite non-trivial to turn on the required discrete flux in a
consistent manner without violating the charge conservation condition
\twostat{}. We will avoid these complications in this paper, by
restricting ourselves to cases where $m_\gamma,n_\gamma$ are even
integers, and by not including any discrete flux at the orientifold
planes.

Finally, $G_{(3)}$ obeys an imaginary self-duality (ISD) condition,
$*_6 G_{(3)} = i G_{(3)}$, as will be shown in the next section.  This
condition implies that the 3-form flux contributes positively to the
total D3-brane charge.  To see this note that the ISD condition
implies that
\eqn\selfd{*_6 H_{(3)}/g_s = - ( F_{(3)} - C_{(0)} H_{(3)}). }
Since $H_{(3)}\wedge F_{(3)} = H_{(3)}\wedge (F_{(3)}-C_{(0)}
H_{(3)})$, we learn that\foot{In the conventions of \GKP, $H_{(3)}
\wedge *_6 H_{(3)} = - {1\over3!} H_{mnp} H^{mnp}\ {\rm
Vol}$, where $m,n,p$ are real coordinates on ${\cal M}_6$ and ${\rm
Vol}$ is the volume form.}
\eqn\selfth{ \eqalign {\int_{{\cal M}_6} H_{(3)} \wedge F_{(3)}=&
 - {1\over g_s}\int_{{\cal M}_6} H_{(3)} \wedge *_6 H_{(3)} \cr &=
{1\over g_s}{1\over3!}\int_{{\cal M}_6} \sqrt{g_{{\cal M}_6}}
\ H_{(3)}{}^2 >0.}}
Therefore, in the presence of nontrivial RR and NS fluxes which carry
nonzero $N_{\rm{flux}}$, the number of D3 branes required to saturate
\twostat{}\ will always be fewer than 16.\foot{We do not allow the presence of
anti D3-branes, since our main interest is SUSY solutions.  Some
aspects of non-supersymmetric vacua with anti D3-branes and fluxes
have recently been described in \KPV}.  In fact, in some models, one
can entirely cancel the tadpole with fluxes.

\subsec{The Scalar Potential from 3-form flux}

\subseclab\tension

Turning on three-form fluxes gives rise to a potential for some of the
moduli.  The four dimensional effective theory has a term of the form
\GKP
\eqn\Lgsquared{{\cal L}_G =
{1\over4\kappa_{10}{}^2}\int_{{\cal M}_6}d^6 y 
{G_{(3)}\wedge *_6 \bar G_{(3)}\over \Im\phi},}
which arises from the $G_{(3)}\cdot\bar G_{(3)}$ term in the ten
dimensional action \IIBsugra. To understand why this term gives rise
to a potential for some moduli it is useful to write
\eqn\Gparts{G_{(3)} = G^{\rm ISD} + G^{\rm IASD},}
where
\eqn\ISDdef{\eqalign{
*_6 G^{\rm ISD}  &= +i G^{\rm ISD},\cr
*_6 G^{\rm IASD} &= -i G^{\rm IASD}.}}
Then,
\eqn\PotAndTension{\eqalign{{\cal L}_G
&= {1\over2\kappa_{10}{}^2\Im\phi}
\int_{{\cal M}_6} G^{\rm IASD}\wedge *_6 \bar G^{\rm IASD}
-{i\over4\kappa_{10}{}^2\Im\phi}
\int_{{\cal M}_6} G_{(3)}\wedge\bar G_{(3)}\cr
&=\qquad\qquad\qquad\qquad
{\cal V}_{\rm scalar}\qquad + \qquad {\rm topological}.}}
The second term in \PotAndTension\ is topological. It is proportional to
$N_{\rm{flux}}$ \twostat{}\ and independent of moduli.  One expects on
general grounds that three-form flux configurations, which give rise
to D3-brane charge, should also lead to D3-brane tension.  This
contribution to D3-brane tension is accounted for by the second term.

The first term in \PotAndTension\ gives rise to the scalar potential
and is central to this paper. It is positive semidefinite and vanishes
when the flux meets the imaginary self-duality condition. The moduli
dependence enters in two ways.  First, $G_{(3)}$ depends on the
axion-dilaton \fielddefs.  Second, the decomposition of $G_{(3)}$ into
ISD and IASD parts, depends on some metric moduli.  Requiring that
$G_{(3)}$ is imaginary self dual fixes many of these moduli.

\subsec{IIB on the orientifold $T^6/Z_2$}
Let us now focus on IIB string theory compactified on a $T^6/Z_2$
orientifold.  The six transverse directions will be denoted as
$x^i,y^i$, $i=1,\ldots,3$.  The orientifold action can be denoted as
$\Omega R (-1)^{F_L}$, where $R$ stands for a reflection of all of the
compactified dimensions $(x^i,y^i) \rightarrow -(x^i,y^i), i=1,
\cdots, 3$.  In fact the model is related to the Type I theory
compactified on $T^6$ by six T-dualities along all the compactified
directions. It preserves ${\cal N}=4$ supersymmetry, i.e., $16$
supercharges.

The massless fields after compactification arise from the massless
fields in the IIB ten dimensional supergravity theory.  The bosonic
fields in the ten dimensional theory are the metric $g_{MN}$, the NS
2-form $B_{(2)}$, the RR fields $C_{(2)}$,$C_{(4)}$, and the
dilaton-axion $\phi,C_{(0)}$.  Their transformation properties under
$\Omega (-1)^{F_L}$ are as follows:
\eqn\transo{\matrix{& \Omega & \ (-1)^{F_L} \cr
            g_{MN}  & + & + \cr
            B_{(2)} & - & + \cr
            C_{(2)} & + & - \cr
            C_{(4)} & - & - \cr
            C_{(0)} & - & - \cr
            \phi    & + & + }}
The resulting massless bosonic fields are then:
\eqn\spec{\matrix{%
   \hfill g_{\mu\nu}       & \hfill  1 & \hbox{graviton}\hfill \cr
   \hfill g_{ab}           & \hfill 21 & \hbox{scalars}\hfill\cr
   \hfill (B_{(2)})_{a\mu} & \hfill  6 & \hbox{gauge bosons}\hfill\cr
   \hfill (C_{(2)})_{a\mu} & \hfill  6 & \hbox{gauge bosons}\hfill\cr
   \hfill (C_{(4)})_{abcd} & \hfill 15 & \hbox{scalars}\hfill\cr
   \hfill C_{(0)}          & \hfill  1 & \hbox{scalar}\hfill\cr
   \hfill \phi             & \hfill  1 & \hbox{scalar}\hfill
}}
We see that the massless fields which survive the orientifold
projection are the graviton, 12 gauge bosons and 38 scalars, plus
their fermionic partners.  These are organized into representations of
${\cal N}=4$ supergravity as follows.  The graviton, six gauge bosons
and the axion-dilaton along with their fermionic partners, lie in a
supergravity multiplet \JoeTwo. In addition there are six vector
multiplets each containing a gauge boson, six scalars and their
fermionic partners.  Thus, in the absence of 3-form flux, the moduli
space of $T^6/Z_2$ compactifications is parametrized by 38 scalars.
When 3-form flux is turned on, some of the scalars from $C_{(4)}$
become charged, which means that they obtain Stuckelberg type kinetic
terms $\sim (\partial_\mu\lambda+mA_\mu)^2$, where $m$ is determined
by the flux.  For generic ${\cal N}=1$ solutions, one can show that
twelve of these scalars are eaten by gauge fields though the Higgs
mechanism.  (See, for example, \bergsh\ or \joefrey\ for related
discussions in somewhat different contexts).  Six of these twelve
scalars are partners of metric K\"ahler moduli which also get heavy.
The remaining three scalars from $C_{(4)}$ pair up with three metric
K\"ahler moduli to form three ${\cal N}=1$ chiral multiplets which
survive in the low-energy theory.

In the T-dual of Type I theory on $T^6$, one would also include 16
D3-branes, each with a worldvolume ${\cal N}=4$ vector multiplet.  We
will ignore any brane worldvolume fields for now, and briefly discuss
the physics on the branes we must introduce in \S9.

We discussed above that turning on fluxes leads to a potential on
moduli space.  It is important to note that although some of the
moduli will gain a mass from this potential, the effective field
theory keeping only the fields \spec\ from the closed string sector
(plus any massless open string fields, if branes are introduced) is
valid.  This is because the masses generated by the flux-induced
potential will scale like $m \sim {\alpha' \over R^{3}}$, where we
have assumed an isotropic torus of size $\sim R$.  The KK modes on the
Calabi-Yau geometry have masses that scale like $m_{KK} \sim {1\over
R}$, so if we work at sufficiently large radius (where our
supergravity considerations are most valid in any case), $m <<
m_{KK}$, and we are justified in truncating to the field theory of the
modes \spec.

It is helpful to regard the torus as a complex manifold and organize
the various moduli accordingly. Nine of the twenty-one scalars that
arise from the ten-dimensional metric correspond to K\"ahler
deformations, while the remaining twelve scalars correspond to complex
structure deformations.

An essential difference between the six-torus and a Calabi-Yau
three-fold is the following. For a generic $CY_3$, Yau's theorem
implies that any complex structure or K\"ahler deformation corresponds
to a nontrivial deformation of the Ricci-flat metric.  This is not
true for the six-torus or the $T^6/Z_2$ case at hand.  In this case,
as we will see below, the complex structure is specified by nine
complex parameters.  Three of these parameters correspond to
deformations of the complex structure at fixed metric.

\subsec{The Complex Structure of a Torus}
Nine complex coordinates are needed to describe the complex structure
of $T^6$.  We will use the explicit parametrization discussed in
\Moore, which is summarized below.  Let $x^i,y^i$, $i = 1,\ldots,3$
be six real coordinates on $T^6$ which are periodic, $x^i\equiv
x^i+1$, $y^i\equiv y^i+1$, and take the holomorphic 1-forms to be
$dz^i=dx^i + \tau^{ij}dy^j$. The complex structure is completely
specified by the period matrix $\tau^{ij}$.  We choose the
orientation\foot{This choice of orientation is different that in
\Moore\ and is chosen to be consistent with the conventions of \GKP.}
\eqn\orientation{\int dx^1\wedge dx^2\wedge dx^3\wedge
                      dy^1\wedge dy^2\wedge dy^3=1,}
and use the following basis of $H^3(T^6,{\bf Z})$:
\eqn\basis{\eqalign{
  \alpha_0 &= dx^1\wedge dx^2\wedge dx^3, \cr
  \alpha_{ij} &= {1\over2}\epsilon_{ilm}
                 dx^l\wedge dx^m\wedge dy^j,\quad 1\le i,j\le3, \cr
  \beta^{ij} &= -{1\over2}\epsilon_{jlm}
                 dy^l\wedge dy^m\wedge dx^i, \quad 1\le i,j\le3, \cr
  \beta_0 &= dy^1\wedge dy^2\wedge dy^3. }
}
This basis satisfies the property
\eqn\symp{\int_{{\cal M}_6} \alpha_I \wedge \beta^J = \delta^J_I.}
Finally, we choose a normalization so that the holomorphic three-form
$\Omega$ is
\eqn\holot{\Omega=dz^1\wedge dz^2\wedge dz^3.}
One can show that
\eqn\altome{\Omega=\alpha_0 + \alpha_{ij}\tau^{ij}  - \beta^{ij}
(\cof\tau)_{ij} + \beta^0(\det\tau),}
where
\eqn\defcof{(\cof\tau)_{ij}\equiv (\det\tau) \tau^{-1,{\rm T}} = {1 \over
2} \epsilon_{ikm} \epsilon_{jpq} \tau^{kp} \tau^{mq}.}

\subsec{The RR and NS flux}
The flux that we turn on must be even under the $Z_2$ orientifold
symmetry.  The intrinsic parity, under $\Omega (-1)^{F_L}$, of the
various fields is given in \transo.  One sees that the 3-form field
strengths $F_{(3)}$, $H_{(3)}$ that are excited must be proportional
to 3-forms of odd intrinsic parity.  However, the quantity that must
be even is the total parity, which for a $p$-form on the internal
space is the product of this intrinsic parity and an explicit $(-1)^p$
from the reflection action on the indices \GKP.  Therefore, the 3-form
field strengths must transform as $(F_{(3)})_{abc} \rightarrow
(F_{(3)})_{abc}$, $(H_{(3)})_{abc}\rightarrow (H_{(3)})_{abc}$ under
the $Z_2$ action.  Similarly, the field strength $F_{(5)}$ must be
proportional to a 5-form of even intrinsic parity.  We will ensure
below that the three-forms which are excited have the correct symmetry
properties.  The resulting 5-form field strength is then determined by
the equations of motion \bianchi, and automatically satisfies the
correct symmetry properties.

Note that the Bianchi identities for $F_{(3)}$ and $H_{(3)}$ require
that they be closed.  They should thus be expressible as a linear
combination of the basis vectors of $H^3(T^6,{\bf Z})$. All the basis
elements, \basis, are three forms of odd parity under the $Z_2$ action
which takes $x^i,y^i \rightarrow -x^i,-y^i$.  So the symmetry
constraint mentioned above is automatically taken care of by
expressing the three-forms in this manner. Finally, taking into
account the quantization conditions
\quantization, $F_{(3)}$ and $H_{(3)}$ can be expressed as
\eqn\expthree{\eqalign{
{1 \over (2\pi)^2 \alpha'}F_{(3)}
  &=a^0\alpha_0+a^{ij}\alpha_{ij}+b_{ij}\beta^{ij} + b_0\beta^0,\cr
{1 \over (2 \pi)^2 \alpha' } H_{(3)}
  &= c^0\alpha_0 + c^{ij}\alpha_{ij} + d_{ij}\beta^{ij} +d_0\beta^0.
  }}
Here $a^0,\alpha_{ij},\beta^{ij},\beta^0$ and $c^0,c^{ij},d_{ij},d_0$
are all integers.  We will search for vacua maintaining the ansatz of
constant fluxes \expthree\ on the $T^6$ throughout the paper.

\newsec{Supersymmetry}

\subsec{Spinor conditions}

In the discussion below our conventions are as follows: The $\gamma_i,
i = 0, \cdots, 9$ matrices are all real and satisfy the algebra
$\{\gamma^i,\gamma^j\}=\eta^{ij}$.  The matrix, $\gamma^0$, is
anti-hermitian and the others are hermitian.  Also,
\eqn\defa{\Gamma^{(4)}\equiv i\gamma^0\gamma^1\gamma^2\gamma^3}
and
\eqn\defb{\Gamma^{(6)}\equiv i \gamma_4 \gamma_5 \gamma_6 \gamma_7
\gamma_8 \gamma_9.}
Both $\Gamma^{(4)}, \Gamma^{(6)}$ are hermitian with eigenvalues $\pm
1$.  For the rest we follow the conventions of \GPOne.  Denote the
spinor $\epsilon$ as
\eqn\spone{\epsilon=\epsilon_L+i \epsilon_R.}
Here, $\epsilon_L$ is a Majorana spinor in ten dimensions.  We can
write
\eqn\sptwo{\epsilon_L=u \otimes \chi + u^* \otimes \chi^*,}
where ${}^*$ denotes complex conjugation, and $\Gamma^{(4)} u = u$,
$\Gamma^{(6)} \chi =-\chi$.  The complex conjugate spinors have
opposite $4$ and $6$ dimensional helicity.

Since we are working on a $T^6/Z_2$ orientifold, the spinor must be
invariant with respect to the $Z_2$ symmetry.  The $Z_2$ action
corresponds to $\Omega R_{456789} (-1)^{F_L}$, where $R_{456789}$
stands for a reflection in the six directions. This means that
\eqn\spthree{\epsilon_R=-\gamma_4 \gamma_5 \gamma_6 \gamma_7 \gamma_8
\gamma_9 \epsilon_L.}
That is
\eqn\spfour{i \epsilon_R=- \Gamma^{(6)} \epsilon_L=u \otimes \chi -
u^* \otimes \chi^*,}
which gives from \spone\
\eqn\spfive{\epsilon=2 u \otimes \chi.}
So, the spinor consistent with the $Z_2$ orientifolding symmetry is of
Type B(ecker).

Now following \GPOne\ we are lead to the conditions
\eqn\spsix{G_{(3)}\chi=0,\ G_{(3)} \chi^*=0,\ {\rm and}
\  G_{(3)} \gamma^{\bar\imath} \chi^*=0,}
where we have introduced complex coordinates such that
\eqn\spseven{\gamma^{\bar\imath} \chi=0.}
The first condition in \spsix\ gives
\eqn\speight{(G_{(3)})_{ijk}=0, (G_{(3)})_{ij}^j =0.}
The second that:
\eqn\spnine{(G_{(3)})_{{\bar\imath} {\bar j} {\bar k}} = 0, 
(G_{(3)})_{{\bar\imath} {\bar j}}^{\bar j} =0,}
note the second condition in \spnine\ kills off the $(1,2)$ terms of
the kind $J\wedge d{\bar z^a}$.  Finally the third condition in
\spsix\ gives:
\eqn\spten{(G_{(3)})_{{\bar\imath } {\bar\jmath} l} =0}
Putting all this together only primitive $(2,1)$ terms in $G_{(3)}$
survive.  Primitivity means that
\eqn\seq{J \wedge G_{(3)}=0.}
For  a $(2,1)$ form this is equivalent to requiring that
\eqn\prim{g^{i {\bar\jmath}} (G_{(3)})_{li{\bar\jmath}}=0.}

We turn next to analyzing the requirement that $G_{(3)}$ is of $(2,1)$
type and then discuss the requirements imposed by primitivity in
\S3.4.

\subsec{$G_{(3)}$ of type (2,1)}

Another way to phrase the condition that $G_{(3)}$ be of type (2,1) is
that the (0,3), (3,0), and (1,2) terms in $G_{(3)}$ must vanish.  We
saw above that the moduli space of complex structures for $T^6$ can be
parametrized by the period matrix $\tau^{ij}$.  One can show that
\eqn\vara{\partial_{\tau^{ij}} \Omega = k_{ij} \Omega + \chi_{ij},}
where $\chi_{ij}, 1 \le i,j \le 3$ are a complete set of $(2,1)$
forms.  The condition that $G_{(3)}$ is of $(2,1)$ type is then
equivalent to requiring that
\eqn\fullcond{\eqalign{
  \int G_{(3)}\wedge \Omega &=0 \cr
  \int {\bar G}_{(3)}\wedge \Omega &=0 \cr
  \int G_{(3)} \wedge {\chi_{ij}} & =0,\quad 1\le i,j\le3.}
}

A convenient way to impose the requirements \fullcond, is by
constructing the superpotential
\eqn\defsuperpota{W=\int G_{(3)} \wedge \Omega.}
From \vara\ we find that
\eqn\varab{\partial_{\tau^{ij}} W= k_{ij} W+ \int G \wedge \chi_{ij}.}
Similarly,
\eqn\vart{\partial_\phi W= -H \wedge \Omega = {1\over(\phi-\bar\phi)}
\int (G_{(3)}-\bar G_{(3)} )\wedge \Omega }
Thus \fullcond\ is equivalent to demanding that
\eqna\finalcond 
$$\eqalignno{
  W &=0 &\finalcond a\cr
  \partial_\phi W & =0 &\finalcond b\cr
  \partial_{\tau^{ij}} W & = 0 &\finalcond c}$$

\subsec{The Superpotential and Equations for SUSY Vacua}

Using \expthree\ it follows that the superpotential \defsuperpota, is:
\eqn\superpotb{{1 \over (2\pi)^2 \alpha'} W =
(a^0-\phi c^0)\det\tau -(a^{ij}-\phi c^{ij}) (\cof\,\tau)_{ij}
- (b_{ij} -\phi d_{ij}) \tau^{ij} - (b_0-\phi d_0).}
We see from \superpotb, that it depends on ten complex
variables---$\phi$ and the nine components of $\tau^{ij}$.  But,
equations \finalcond{}\ give rise to eleven equations in these
variables. Thus, generically all the equations
\finalcond{a\hbox{--}c}\ cannot be met and supersymmetry is broken.

The explicit equations of motion that follow from \finalcond{}\ and
\superpotb\ are
\eqna\susyeqs
$$\eqalignno{a^0\det\tau - a^{ij}(\cof\tau)_{ij}
- b_{ij}\tau^{ij} - b_0 &= 0,&\susyeqs a\cr
c^0\det\tau - c^{ij}(\cof\tau)_{ij}
- d_{ij}\tau^{ij} - d^0 &= 0,&\susyeqs b}$$
$$(a^0-\phi\,c^0)(\cof\tau)_{kl}
- (a^{ij}-\phi\,c^{ij})\epsilon_{ikm}\epsilon_{jln}\tau^{mn}
- (b_{ij}-\phi\,d_{ij})\delta^i_k\delta^j_l = 0.\eqno\susyeqs c$$
Here, the first equation comes from \finalcond{a}\ minus
\finalcond{b}, the second, from \finalcond{b}, and the third from
\finalcond{c}.\foot{In deriving the third equation, it is
useful to note the relations $\det\tau = {\textstyle{1\over3}}
\epsilon_{ikm}\epsilon_{jln}\tau^{ij}\tau^{kl}\tau^{mn}$, 
and $(\cof\tau)_{ij}
= \half\epsilon_{ikm}\epsilon_{jln}\tau^{kl}\tau^{mn}$}
The equations \susyeqs{} are coupled non-linear equations in several variables
and are difficult to solve in full generality. 

It might seem odd at first glance that all nine scalars parametrizing
the complex structure can be fixed, even though, as was argued in
section 2.3, only six of them correspond to components of the metric
and enter in the supergravity equations of motion.  This happens
because the requirements for ${\cal N}=1$ supersymmetric solutions are
stronger than the requirements which would follow from searching for
generic solutions to the equations of motion.

\subsec{Primitivity }

Once the complex structure is chosen such that $G_{(3)}$ is of $(2,1)$
type, \prim, imposes the requirement of primitivity.  Note that in
\prim\ the index $l$ can take values $\{ 1,2,3 \}$, so primitivity
gives rise to three complex equations or equivalently six real
equations.  The space of K\"ahler forms is 9 dimensional to begin with
so generically this will leave a three dimensional moduli space of
K\"ahler deformations\foot{The surviving K\"ahler moduli have axionic
partners which come from the $C_{4}$ field, together these give rise
to three chiral superfields at low energies. The six K\"ahler moduli
which get heavy also have partners, these obtain a mass due to
Chern-Simons couplings \IIBsugra, \Ffive.}.

Equation \prim\ can be thought of as $6$ linear equations in the $9$
metric components $g^{i {\bar\jmath}}$.  Solving these is relatively
straightforward. In contrast we saw above that requiring $G$ to be of
type $(2,1)$ results in coupled non-linear equations which are
considerably harder to work with.  In practice, in the examples below,
it will sometimes be easier to ensure primitivity by directly imposing
the condition \seq\ on the K\"ahler two-form.

It is worth making one more comment at this stage.  We mentioned in
section 2.1 that the equations of motion can be solved if $G_{(3)}$ is
an imaginary self-dual three form.  This allows $G_{(3)}$ to be of
three types: primitive $(2,1)$, $(0,3)$, or $(1,2)$ of the kind $J
\wedge d{\bar z}^a$. We also saw in section 2.2 that in all these
cases, the scalar potential for the moduli was minimized and equal to
zero.  Supersymmetry on the other hand is preserved if $G_{(3)}$ is
purely a primitive $(2,1)$ form. Thus for choices of complex structure
and K\"ahler class where $G_{(3)}$ has $(0,3)$ or $(1,2)$ terms, some
auxilary $F$ or $D$ term must get a vev. However, since the potential
continues to vanish in these cases, these $F$- and $D$-terms cannot be
present in the scalar potential. Part of this discussion is already
familiar from the study of a generic Calabi Yau manifold \GKP. If
$G_{(3)}$ has a $(0,3)$ term the F-component of the volume modulus
gets a vacuum expectation value, however this F-component does not
enter the potential because of the no-scale structure of the
four-dimensional supergravity theory.  Similarly when $(1,2)$ terms are
present auxiliary D-terms must acquire expectation values in general.
The absence of these terms in the potential can probably best be
understood in the context of the underlying ${\cal N}=4$ supersymmetry
present in the $T^6/Z_2$ case.  We leave a more systematic analysis of
the low-energy supergravity theory along the lines of
\refs{\louishet,\kalmy,\bergsh} for future work; such analyses for the
case of generic Calabi-Yau threefolds with fluxes have appeared in
e.g. \refs{\mich,\dallgata,\ferrara}.

\newsec{Some Supersymmetric Solutions}

The equations which determine the value of the moduli are dificult to
solve in general. The main challenge are the coupled non-linear
equations \susyeqs{}\ which determine the complex structure of the
torus.

We do not solve these equations in their full generality
below. Instead in section 4.1 we discuss some examples, where the
fluxes take simple values that allow for analytic solutions.  Already
these simpler cases are quite interesting.  As we will see, in many
cases, stable minima exist where all the complex structure moduli and
some of the K\"ahler moduli are stabilized. Section 4.2 deals with the
inverse problem: we start with some values for the moduli and ask for
fluxes which stabilize the moduli at these values consistent with
supersymmetry.  The inverse problem is sometimes easier to solve. The
solutions in section 4.1 have $\ncal =1$ supersymmetry. With a few
possible exceptions this should be true of the vacua in section 4.2 as
well.  Section 4.3 analyses some additional cases where the fluxes
lead to tractable solutions. These examples turn out to have $\ncal
=3$ supersymmetry.  Finally, some comments related to obtaining a
general supersymmetric solution are in section 4.4.

Not all of the solutions studied in this section are physically
distinct. Section 5 discusses how solutions related by $SL(2,{\bf Z})
\times SL(6,{\bf Z})$ transformations should be identified. Starting
with some of solutions found in this section, other physically
distinct solutions can be obtained by rescaling the fluxes, or
carrying out $GL(2,{\bf Z}) \times GL(6,{\bf Z})$
transformations. This is illustrated in some examples here and
discussed more fully in section 6.

One final comment before turning to examples.  One would like to know
if the analysis of ${\cal N}=1$ supersymmetric vacua in this section,
receives significant $\alpha'$ and $g_s$ corrections.  We have not
discussed an explicit ${\cal N}=1$ superspace description of the the
low-energy effective theory in the presence of fluxes in this
paper. But it is clear that such a description would involve both a
superpotential \defsuperpota, and $D$-terms \foot{These play a role in
ensuring primitivity of $G_{(3)}$ for example.}. The superpotential
must be exact in the $\alpha'$ expansion since the partner of volume
modulus is an axion which cannot occur in the $\alpha'$ (or string
loop) corrections to the superpotential.  Quite plausibly, in this
case, this is true of the $D$ terms as well, since they are related by
the underlying ${\cal N}=4$ symmetry to the $F$-terms.  The dilaton in
the examples below is typically stabilized at a value of order one.
One can be hopeful that the resulting $g_s$ corrections (e.g. to the
K\"ahler potential of the low-energy field theory) do not
qualitatively alter our conclusions, at least in some of the examples
studied here.

\subsec{Example 1: Fluxes proportional to the identity}

We begin by studying the case where,   
\eqn\identfluxes{(a^{ij},b_{ij},c^{ij},d_{ij}) =
(a,b,c,d)\,\delta_{ij},}
that is all the flux matrices are diagonal and proportional to the
identity.

The equations determining the complex structure, \susyeqs{}\ will be
considered first, followed by the conditions for primitivity.

With the flux matrices of the form \identfluxes, it is easy to see
from \susyeqs{}, that the period matrix must be diagonal,
\eqn\diagp{ \tau^{ij}=\tau\delta^{ij}.}
(In fact this is more generally true if the flux matrices are all
diagonal).

The equations of motion \susyeqs{}\ then take the form 
\eqn\pone{P_1(\tau)\equiv a^0\tau^3 - 3a\tau^2 - 3b\tau - b_0 = 0,}
\eqn\ptwo{P_2(\tau)\equiv c^0\tau^3 - 3c\tau^2 - 3d\tau - d_0 = 0,}
\eqn\pthree{(a^0-\phi\,c^0)\tau^2 - 2(a-\phi\,c)\tau - (b-\phi\,d) = 0.}
We are only interested in solutions in which $\tau$ is complex (since
solutions with $\Im(\tau)=0$ lie at boundaries of the moduli
space). It is straightforward to show that in this case\foot{$P_1$ and
$P_2$ are cubic polynomials with real coefficients, that share a
common complex root, $\tau$.  Therefore, $\bar\tau$ is also a root,
and the two equations share a common quadratic factor.  This common
factor is proportional to $P = c^0P_1-a^0P_2$, which has integer
coefficients.  Since $P_1$ and $P_2$ also have integer coefficients,
it follows that $P_1/P$ and $P_2/P$ are each binomials with rational
coefficients. But, a polynomial with integer coefficients that
factorizes over the rationals also factorizes over the integers.},
\eqna\pfour
$$P_1(\tau) = (f\tau+g)P(\tau),
\quad 
P_2(\tau) = (h\tau+k)P(\tau),\eqno\pfour a$$
for some
$$P(\tau) = l\tau^2+m\tau+n,\quad f,g,h,k,l,m,n\in{\bf Z}.
\eqno\pfour b$$
Thus, $\tau$ is a root of $P(\tau)$ and $\phi$ is determined from
equation \pthree.  Note that not every septuple $(f,g,h,k,l,m,n)$
corresponds to integral flux.  From the relations
\eqn\pseven{\eqalign{fm+gl&=-3a,\cr
                    fn+gm&=-3b,}\quad
           \eqalign{hm+kl&=-3c,\cr
                    hn+km&=-3d,}}
we have consistency conditions modulo 3.

The D3-brane charge of the flux in this solution is given by
\eqn\peight{\eqalign{N_{\rm flux} = 
{1\over(2\pi)^4(\alpha')^2}\int H_{(3)}\wedge F_{(3)}
& = \bigl(b_0c^0-a^0d_0\bigr)+3(bc-ad)\cr
& = - {1\over3}(fk-gh)(m^2-4ln),}}
which has the property that it is always $0\,(\mod\,3)$.\foot{To see
this, note that \pseven\ can be written as $({f\atop h}{g\atop k})
({m\atop l}{n\atop m}) = -3 ({a\atop c}{b\atop d})$.  Since $({m\atop
l}{n\atop m}) \equiv ({m\atop -2l}{-2n\atop m})\ (\mod\,3)$, this
means that $({f\atop h}{g\atop k}) ({m\atop -2l}{-2n\atop m}) \equiv
0\ (\mod\,3)$.  Taking the determinant of both sides then gives
$(fk-gh) (m^2-4ln) \equiv 0\ (\mod\,9).$} One can also show that the
result \peight\ is explicitly positive in our conventions.\foot{ Our
conventions are $\Im\tau, \Im\phi > 0$.  One can show that the factor
$(fk-gh)$ in \peight\ satisfies ${\rm sign}(fk-gh) = {\rm sign}
(\Im\phi/\Im\tau)$. Therefore it is positive.  The other factor,
$(m^2-4ln)$, is the discriminant of $P(\tau)$.  It is negative since
the roots are complex.}

In summary, starting with fluxes of the form \identfluxes, the
neccessary and sufficient condition for a non-singular solution, is
the existence of integers $(f,g,h,k,l,m,n)$ which satisfy the
conditions, \pseven, and which give rise to nonzero three brane
charge, \peight.

In practice, determining polynomials of the form \pfour{}, by direct
scrutiny is often easier than finding appropriate septuples
$(f,g,h,k,l,m,n)$.

As a concrete example, consider the case
\eqn\psone{P_1(\tau)\equiv \tau^3-1=0}
\eqn\pstwo{P_2(\tau)\equiv \tau^3 + 3 \tau^2 + 3 \tau + 2=0}

Both polynomials share a common factor $P(\tau)=\tau^2+\tau+1$
and can be expressed as:
\eqn\psten{P_1\equiv (\tau-1)P(\tau)=0}
\eqn\pseleven{P_2\equiv (\tau+2)P(\tau)=0.}
Solving  $P(\tau)=0$ with the condition
$\Im(\tau)>0$,  gives
\eqn\valtau{\tau=e^{2\pi i \over 3}.}
$\phi$ is obtained from \pthree, and given by
\eqn\valphi{\phi=\tau=e^{2 \pi i\over 3}.}

We see that the moduli are fixed at a very symmetric point.  Since the
period matrix is diagonal, the torus factorizes as $T^6 \equiv T^2
\times T^2 \times T^2$ with respect to complex structure. In
fact, when viewed in F-theory, this factorization becomes $T^8 \equiv
T^2 \times T^2 \times T^2 \times T^2$. Since the eigenvalues of the
period matrix are all equal to one another, and to value of the
dilaton-axion, all the four 2-tori have the same modular parameter.

From \psten, \pseleven, we see that the septuple 
\eqn\valsep{(f,g,h,k,l,m,n) =(1,-1,1,2,1,1,1).}
Also from \psone, \pstwo, and \pone, \ptwo, we see that the integers
\eqn\pstwelve{(a^0,a, b,b_0) = (1,0,0,1)
\quad (c^0,c,d,d_0) = (1,-1,-1,-2)}
Either way, we find that the three-brane charge carried by the flux is
given by
\eqn\psixteen{N_{\rm flux}= {1\over(2\pi)^4(\alpha')^2}
\int H_{(3)}\wedge F_{(3)}=3.}

Notice that most of the non-zero fluxes in \pstwelve\ are odd integer.
We discussed in Section 2.1 why consistency on the $T^6/Z_2$
orientifold requires additional discrete flux to be turned on when odd
integer flux is present.

To avoid this complication we can simply choose the fluxes to be twice
the values indicated in \pstwelve. That is
\eqn\psthirteen{(a^0,a, b,b_0) = (2,0,0,2)
\quad (c^0,c,d,d_0) = (2,-2,-2,-4),}
and,
\eqn\psfourteen{(f,g,h,k,l,m,n)=(2,-2,2,4,1,1,1).}
No discrete flux in needed now.  Since doubling all fluxes simply
rescales the superpotential by an overall factor, the equations
determining the moduli \susyeqs{}, remain the same and therefore the
solutions for the moduli are still given by \valtau, \valphi.

From \peight,  we see that after doubling the fluxes 
\eqn\finalnf{N_{\rm flux}= 12.}
Eq. \twostat{}, now implies that for a consistent solution we need to
add ten wandering branes in addition, i.e., $N_{D3}=10$.

This completes our discussion of how the complex structure moduli are
determined, in this case. To complete our analysis we must next impose
the requirement that the three flux $G_{(3)}$ is primitive. Before
doing so though, let us pause to make two comments.

First, other closely related examples can be obtained by starting with
the fluxes \pstwelve, and doing other rescalings.  For example, one
can double the $H_{(3)}$ flux while increasing the $F_{(3)}$ flux by a
factor of four so that the resulting values for the fluxes are:
\eqn\forty{(a^0,a, b,b_0) = (4,0,0,4)
\quad (c^0,c,d,d_0) = (2,-2,-2,-4).}
Now, it is straightforward to see from \pone, \ptwo, that the
resulting value for $\tau$, $\phi$, are:
\eqn\valtwo{ \tau=e^{2 \pi i \over 3}, \quad \phi=2 e^{2 \pi i \over 3}. }
The resulting contribution to three brane charge is given by:
\eqn\vvflu{N_{\rm flux}= 24,}
so that the $N_{D3}=4$. The rescalings discussed in \forty, illustrate
a more general feature which will be dealt with in more generality in
section 6: given a solution, additional ones can be obtained by
carrying out $GL(6,{\bf Z}) \times GL(2,{\bf Z})$ transformations on
the fluxes and the moduli, provided the resulting contribution to
D3-brane charge is within bounds.

Second, one would like to know whether there are other solutions with
fluxes of the form, \identfluxes, which are not related to those
discussed above by $GL(6,{\bf Z}) \times GL(2, {\bf Z})$
transformations or rescalings.  While we do not give all the details
here, it is straightforward to tabulate all choices of fluxes (or
equivalently choices of the septuple $(f,g,h,k,l,m,n)$) which meet the
requirements for the existence of ${\cal N}=1$ supersymmetric
solutions.  In all these cases one finds that the resulting values for
the moduli are related to \valtau, \valphi, by a rescaling or
$GL(6,{\bf Z}) \times GL(2, {\bf Z})$ transformations. We have not
studied the corresponding fluxes exhaustively, but in several cases
they too are related to \pstwelve, by the same rescaling or $GL(6,{\bf
Z}) \times GL(2,{\bf Z})$ transformation.

\medskip
\noindent{\it{Primitivity}}
\medskip

We must also verify that (at least on some subspace of the K\"ahler
moduli space), the $G_{(3)}$ flux found from the superpotential above
is primitive.  We will go through this for the flux in Example 1.  A
similar analysis (without substantially more complexity) would apply
to our other examples.

In the case at hand, the flux takes the form \identfluxes. 
More explicitly,
\eqn\exoFandG{\eqalign{%
F & = a^0 dx^1\wedge dx^2\wedge dx^3
    + a \big(dx^1\wedge dx^2\wedge dy^3 +\hbox{cyc.\ perms of 123}\big),\cr
&-b \big(dx^1 \wedge dy^2 \wedge dy^3  +\hbox{ cyc.\ perms of 123}\big)
+b_0 dy^1 \wedge dy^2 \wedge dy^3, \cr
H & = c^0 dx^1\wedge dx^2\wedge dx^3
 + c \big(dx^1\wedge dx^2\wedge dy^3 +\hbox{ cyc.\ perms of 123}\big),\cr
&     -d \big(dx^1\wedge dy^2\wedge dy^3 +\hbox{ cyc.\ perms of 123}\big)
     + d_0 dy^1 \wedge dy^2 \wedge dy^3.
}}


In the present example, it is convenient to impose the requirement of
primitivity in the form of \seq,
\eqn\prim{J \wedge G_{(3)} = 0.}
We are interested in the subspace of K\"ahler forms for which this
requirement is met.

Take $J$ to be of the form
\eqn\jansatz{J = \sum_{a=1}^{3} r_{a}^2 ~dz^a \wedge d\bar z^a
\sim \sum_{a=1}^{3} i r_a^2 ~dx_a \wedge dy_a}
where the second expression uses the fact that the complex structure
$\tau$ of all the three $T^2$s, as given in \valtau, are equal.  Now,
notice that each term in $F$ and $H$ as given in \exoFandG\ contains
no repeat superscripts: one either chooses $dx^a$ or $dy^a$ for each
of $a=1,2,3$, and then wedges the three one-forms together.  But the
K\"ahler form in \jansatz\ contains a sum of two-forms, each of which
looks like $dx^a \wedge dy^a$.  The wedge product of each such term
with $G_{(3)}$ will clearly vanish, because it hits either another
$dx^a$ or another $dy^a$ in each term in $F$ and $H$.  Therefore,
$J\wedge G_{(3)} = 0$ for the most general $J$ of the form \jansatz.

Is there a larger subspace of K\"ahler moduli space that preserves the
primitivity?  Since $G$ is of type (2,1) and J is a (1,1) form, $J
\wedge G$ is a (3,2) form.  There are three nontrivial (3,2) forms on
the $T^6$, so we expect that requiring $J \wedge G = 0$ will yield
three nontrivial complex equations.  The space of K\"ahler forms has
real dimension 9, so generically we expect only a three-dimensional
subspace of the K\"ahler moduli space (suitably complexified by the
addition of axions in the relevant chiral multiplets) to parametrize
flat directions of this ${\cal N}=1$ theory.  However, in the case at
hand, the $G_{(3)}$ flux is particularly simple and non-generic, and
the number of flat directions parametrized by K\"ahler moduli is 6
instead of 3.  One can see the three ``extra'' flat directions by
inspection.  For instance, consider the two-form
\eqn\examp{\omega \sim i (dx^1 \wedge dy^2 + dx^2 \wedge dy^1)}
One can easily check from \exoFandG\ that $\omega \wedge G = 0$.
Further, since the complex structure of all three $T^2$'s is the same,
it is easy to check that
\eqn\examp{\omega \sim dz^1\wedge d{\bar z}^2 + dz^2\wedge d{\bar z}^1,}
so that $\omega$ is of type $(1,1)$.  Analogous perturbations with
$\{1,2\}$ replaced by $\{1,3\}$ and $\{2,3\}$ similarly maintain the
primitivity of $G_{(3)}$.  So the ${\cal N}=1$ vacua persist along a
six-dimensional slice of the K\"ahler moduli space.

One final comment is in order.  Our analysis has ensured that the
solutions discussed above have at least $\ncal = 1$ supersymmetry, but
it does not preclude the possibility of enhanced supersymmetry.  A
simple check is the following: enhanced supersymmetry requires that
additional choices of complex structure are possible, in which
$G_{(3)}$ is still of the kind $(2,1)$ (and primitive).  $\ncal =2$
and $\ncal=3$ require one and two additional choices of complex
structure respectively.  In the solutions above, with $T^6 \equiv
T^2\times T^2\times T^2$, there is a complete permutation symmetry
among the three two-tori.  This ensures that, upto an overall
constant, $G_{(3)}$ must have the form,
\eqn\symmonea{G_{(3)} \sim (dz^1\wedge dz^2\wedge d{\bar z}^3 
+ dz^2 \wedge dz^3 \wedge d{\bar z}^1 
+ dz^3 \wedge dz^1 \wedge d{\bar z}^2).}
Other choices of complex structure can be made, by taking $z^i
\rightarrow {\bar z}^i$ for some or all of the three $T^2$'s, but none
of them preserve the $(2,1)$ nature of $G_{(3)}$.  So we see that
these examples have only $\ncal=1$ supersymmetry.  A detailed
examination of the conditions for $\ncal=2$ supersymmetry is presented
in Section 7, and some more comments on this matter can be found
there.

\subsec{The inverse problem: fluxes from moduli}
\def\atilde{\tilde a}
\def\btilde{\tilde b}
\def\ctilde{\tilde c}
\def\dtilde{\tilde d}

In the previous section we started with some fluxes and asked what are
the resulting values for moduli in an ${\cal N}=1$ susy vacuum.  In
this section we address the inverse problem, namely: start with some
values for the moduli and ask if there are fluxes which can be turned
on such that the resulting potential stabilizes the moduli at the
values we begin with, while preserving ${\cal N}=1$ susy.  The inverse
problem is sometimes easier to solve and helpful in understanding the
full set of consistent vacuua.

Our discussion will not be exhaustive. Instead we will consider one
illustrative case. In section 4.1 we started with flux matrices which
were all proportional to the identity \identfluxes, then argued that
the period matrix must be a multiple of the identity.  Here, we start
by fixing the period matrix to be a multiple of the identity:
\eqn\inperiod{\tau^{ij}= {\rm diag}(\tau,\tau,\tau),}
then ask what values of the fluxes can yield such a solution while
preserving ${\cal N}=1$ supersymmetry.  Our notation in this section
will be chosen to be consistent with Section 4.1.

We begin by writing 
\eqn\IVone{a^{ij} = a\delta^{ij} + \atilde^{ij},\ \ \tr\,\atilde = 0,}
with similar relations for $b,c,d$.  Equations \susyeqs{a}\ and
\susyeqs{b}\ then become
\eqn\IVtwo{a^0\tau^3 - 3a\tau^2 - 3b\tau - b_0 = 0,}
\eqn\IVthree{c^0\tau^3 - 3c\tau^2 - 3d\tau - d_0 = 0,}
and $\partial_{\tau^{ij}}W=0$ becomes
\eqn\IVfour{\eqalign{
&(a^0-\phi\,c^0)\tau^2 - 2(a-\phi\,c)\tau - (b-\phi\,d) = 0,\cr
&(\atilde^{ji}-\phi\,\ctilde^{ji})\tau -
(\btilde_{ij}-\phi\,\dtilde_{ij}) = 0.}}
Eq. \IVfour\ arises by taking the trace and traceless parts of the
third equation in \susyeqs{}.  It can be be summarized as
\eqn\IVfive{\phi = {a^0\tau^2-2a\tau-b \over c^0\tau^2-2c\tau-d}
= {\atilde^{ji}\tau-\btilde_{ij} \over \ctilde^{ji}\tau-\dtilde_{ij}}\,.}
In the notation of \pone, \ptwo, and \pfour{}, the first expression
for $\phi$ in \IVfive\ is
\eqn\IVsix{{P_1(\tau)\over P_2(\tau)} 
= {\bigl((f\tau+g)P(\tau)\bigr)'\over \bigl((h\tau+k)P(\tau)\bigr)'},}
where a prime denotes differentiation with respect to $\tau$.  At
$P(\tau)=0$, this reduces to $(f\tau+g)/(h\tau+k)$ and \IVfive\
becomes
\eqn\IVseven{\phi
= {f\tau+g\over h\tau+k}
= {\atilde^{ji}\tau-\btilde_{ij} \over \ctilde^{ji}\tau-\dtilde_{ij}}\,.}
So, given a solution with $a^{ij},b_{ij},c^{ij},d_{ij}$ proportional
to the identity, we can generate a new solution with the same $\tau$
by, for example, turning on
\eqn\IVeight{
\atilde^{ji} = f n_{ij},\quad
\btilde^{ji} = -g n_{ij},\quad
\ctilde^{ij} = h n_{ij},\quad
\dtilde^{ij} = -k n_{ij},}
with $n_{ij}$ an arbitrary traceless integer-valued matrix.  This is
still not the most general solution.  For each $i,j$, equation
\IVseven\ is two real equations in the four integers
$\atilde^{ji},\btilde_{ij},\ctilde^{ji},\dtilde_{ij}$, for which we
have found a ${\bf Z}$'s worth of solutions parametrized by $n_{ij}
\in {\bf Z}$.  More complicated solutions will fill out a ${\bf Z}^2$'s
worth for each $i,j$.  In addition, the requirement that, for example,
$a$ and $\atilde^{ij}$ each be integer valued is too strict.  We really
only require $a\delta^{ij}+\atilde^{ij}=a^{ij}$ to be integer valued,
and similarly for $b,c,d$.

Finally, the D3-brane charge from flux in this solution can be shown
to generalize from \peight\ to
\eqn\IVfluxcharge{N_{\rm flux} =
{1\over(2\pi)^4(\alpha')^2}\int H_{(3)}\wedge F_{(3)}
= - {1\over3}\bigl(1+{1\over3}\sum_{ij}n_{ij}{}^2\bigr)
(fk-gh)(m^2-4ln).}

As a concrete example consider starting with the values:
\eqn\invfluxv{(a^0,a,b,b_0)=(2,0,0,2), \quad  (c^0,c,d,d_0)=(2,-2,-2,4),}
which were considered in \psthirteen, of section 4.1.
In this case,
\eqn\invsept{(f,g,h,k,l,m,n)=(2,-2,2,4,1,1,1).} 
Since \IVtwo\ and \IVthree, are the same as \pone and \ptwo, $\tau$ is
given by \valtau. Also, since the first equation in \IVfour, is the
same as \pthree, $\phi$ is given by $\valphi$.

The D3-brane charge is given by, \IVfluxcharge,
\eqn\invalflux{N_{\rm{flux}}=12 + 4 \sum_{ij}n_{ij}^2.}
Now it is easy to find many non-diagonal matrices where $\sum_{ij}
n_{ij}^2=1,2,3,4,5$.  Each of them gives a consistent solution, with
$N_{\rm{flux}}$ taking values, $N_{\rm{flux}}=12,16,20,24,28,32$
respectively. Also, we should point out that since, $(f,g,h,k)$ are
even \invsept, the resulting values of ${\tilde a}^{ij}, {\tilde
b}_{ij}, {\tilde c}^{ij},{\tilde d}_{ij}$ are all even as well,
\IVeight, and thus all the fluxes are even.

One last comment.  We argued towards the end of the previous section
4.1 that the examples discussed in it had $\ncal=1$ supersymmetry, and
no more. The examples in this section are closely related to those in
section 4.1, and we expect that they too will generically have only
$\ncal=1$ supersymmetry.

\subsec{More general fluxes}
Section 4.1, discussed the case where the flux matrices $(a^{ij},
b_{ij}, c^{ij},d_{ij})$ are proportional to the identity matrix. Here
we would like to consider flux matrices which are diagonal but with
unequal eigenvalues.  In these cases one can still argue that the
period matrix is diagonal, $\tau^{ij}= {\rm diag}
(\tau_1,\tau_2,\tau_3)$.  As viewed from F-theory then, the resulting
compactification is a product of four two-tori, but the modular
parameters are in general unequal. Unfortunately, solving the
equations for the most general set of diagonal flux matrices is a
difficult task.

To proceed we need to  place additional restrictions on the flux matrices.
Let us begin by considering fluxes of the form:
\eqn\enflux{\eqalign{
\eqalign{a^{ij} &={\rm \diag} (a_1,a_2,a_2),\cr
         b_{ij} &= {\rm diag} (b_1,b_2,b_2),}&\quad
\eqalign{c^{ij} &= {\rm diag} (a^0,c,c),\cr
         d_{ij} &=-{\rm diag}(d_1,a_2,a_2),}\cr
d_0 &= -b_1}}
Setting $\tau^{ij}= {\rm diag}(\tau_1,\tau_2,\tau_3)$,
the superpotential \superpotb, is now given by
\eqn\ensuperpot{\eqalign{
W=& -c^0\phi \tau_1\tau_2\tau_3 + a^0(\tau_1 + \phi)\tau_2\tau_3 
+ c (\tau_2+\tau_3) \phi \tau_1\cr
&-a_1 \tau_2\tau_3 -d_1 \phi \tau_1 -a_2(\tau_1 + \phi) (\tau_2+\tau_3) \cr
&-b_1(\phi+\tau_1) -b_2(\tau_2+\tau_3) -b_0
}}
One can see that the superpotential is symmetric between $\phi
\leftrightarrow \tau_1$ and $\tau_2 \leftrightarrow \tau_3$.  Thus,
for the restricted choice, \enflux, one can consistently seek
solutions where the four modular parameters take at most two distinct
values.

We now turn to describing two examples where additional restrictions
lead to tractable solutions.

\medskip
\noindent{\it Example 2}
\medskip
In the first example, we set
all trilinear and linear terms in the
superpotential, \ensuperpot, to be zero, i.e.,
\eqn\restr{a^0=c=b_1=b_2=0.}
In this case the superpotential takes the form
\eqn\fisup{W=-c^0 \phi \tau_1\tau_2\tau_3 -a_1 \tau_2\tau_3  
- d_1 \phi \tau_1 - a_2(\tau_1 + \phi) (\tau_2+\tau_3) -b_0.}

Setting $\partial_{\phi} W=0, \partial_{\tau_i}W=0$
shows that
\eqn\uneqeq{\tau_1=\phi, \quad  \tau_2=\tau_3,}
(as expected) and in addtion leads to two equations:
\eqn\unea{-c^0\tau_1\tau_2^2-d_1 \tau_1 -2 a_2 \tau_2 =0,}
\eqn\uneb{-c^0\tau_1^2\tau_2-a_1 \tau_2 -2 a_2 \tau_1=0,}
where in both equations we have substituted for $\phi,\tau_3,$ using
\uneqeq.

These lead to the relation,
\eqn\uneqc{\tau_2^2={d_1 \over a_1} \tau_1^2,}
i.e.,
\eqn\uneqd{\tau_2=\pm \sqrt{d_1 \over a_1} \tau_1.}
Substituting in \unea\ gives
\eqn\uneqf{\tau_1=i \sqrt{a_1\over d_1 c^0}\Biggl(d_1\pm 2 a_2 
\sqrt{d_1 \over a_1}\,\Biggr)^{1/2}.}
Setting $W=0$ then leads to a condition determining $b_0$ in terms of
the other flux integers,
\eqn\uneqg{b_0={a_1 \over d_1 c^0}
\Biggl(d_1\pm 2 a_2 \sqrt{d_1 \over a_1}\,\Biggr)^{2}.}
Finally, the contribution to the three brane charge is then given by
\eqn\uneqflux{N_{\rm{flux}}=2 a_1 d_1 + 6 a_2^2 \pm 4 a_2 \sqrt{a_1d_1}.}

To find a consistent non-singular  solution we need to choose integers 
$c^0,a_1,d_1$ and $a_2$ such that
$\tau_1,\tau_2$ are complex, $b_0$ is an integer, and
the total flux $N_{\rm{flux}}$
is within bounds.

One solution to these conditions is obtained by taking
\eqn\uneqvalf{(a_1,d_1,a_2,c^0)=(1,1,-1,-1),}
and choosing the positive sign in \uneqd, so that
\eqn\uneqfinaltc{\tau_2=+\sqrt{d_1\over a_1} \tau_1 =\tau_1.}
Then from \uneqf, we find
\eqn\uneqvalta{\tau_1=i}
and from \uneqflux,
\eqn\uneqvalflux{N_{\rm{flux}}=4.}
Also, from \uneqg, $b_0=-1$ and is indeed an integer.

Notice that the integers \uneqvalf\ are odd. As in example 1, Section
4.1, to avoid complications related to adding discrete flux we can
obtain a consistent solution by doubling all the fluxes so that
\eqn\uneqdo{(a_1,d_1,a_2,c^0,b^0)=(2,2,-2,-2,-2).}
The modular parameters are unchanged and given by \uneqvalta,
\uneqfinaltc, \uneqeq. The total flux is
\eqn\uneqfinalflux{N_{\rm{flux}}=16,}
which means $8$ dynamical D3-branes need to be added for a consistent
solution.

It turns out that the solution above has $\ncal=3$
supersymmetry. Vacua with $\ncal=3$ are analysed in generality in the
recent paper \joefrey.  The possibility of $\ncal=3$ supersymmetry was
also mentioned in \DRS.  The solution above is in fact a special case
of the examples found in \joefrey. To see that it has $\ncal=3$
supersymmetry, we note that with the flux \uneqdo\ and the moduli,
$\phi=\tau^i=i$, $G_{(3)}$ takes the form
\eqn\comthree{{1\over {(2\pi)^2 \alpha'}}
G_{(3)}=2i d{\bar z}^1\wedge dz^2\wedge dz^3.}
It then follows that two additional complex structures in which
$G_{(3)}$ is still of type (2,1) can be defined by taking the complex
coordinates on the three $T^2$'s to be $(w^1,w^2,w^3)= ({\bar z}^1,
{\bar z}^2, z^3) $ or $(w^1,w^2,w^3)=({\bar z}^1,z^2,{\bar z}^3)$.
Thus, as per the general discussion in \joefrey (see also section 4.1
above), the solution has $\ncal=3$ supersymmetry.

Let us also add that additional solutions can be obtained by starting
with the \uneqvalf, \uneqfinaltc, \uneqvalta, and doing $GL(6,{\bf Z})
\times GL(2,{\bf Z})$ transformations.  In particular one can obtain a
solution in which $N_{\rm{flux}}=32$, as will be discussed in more
detail in the examples of section 6.

\medskip
\noindent{\it Example 3}
\medskip
In the next example we again start with flux matrices and
superpotential of the form \enflux, \ensuperpot, respectively, but now
set the following additional restrictions on the fluxes:
\eqn\threeaa{c^0=0, c=-a^0, a_2=0, d_1=-a_1,  b_2=-b_1.}
The superpotential \ensuperpot\ then takes the form
\eqn\threesuper{\eqalign{W= & +a^0(\tau_1 + \phi)\tau_2\tau_3 
-a^0 (\tau_2+\tau_3) \phi \tau_1\cr
&-a_1 \tau_2\tau_3 +a_1 \phi \tau_1  \cr
&-b_1(\phi+\tau_1) +b_1(\tau_2+\tau_3) -b_0.
}}

Solving the equations $\partial_\phi W= 0, \partial_{\tau_i} W=0$, it
is easy to see that
\eqn\threec{\tau_1=\phi=\tau_2=\tau_3 \equiv \tau,}
with $\tau$ given by
\eqn\threevalt{\tau= {a_1 \pm \sqrt{a_1^2-4a^0b_1} \over 2 a^0},}
is a solution.
Setting $W=0$ yields the condition that
\eqn\threebz{b_0=0.}
Finally the D3-brane charge contribution is
\eqn\threeflux{N_{\rm{flux}}=4 b_1a^0-a_1^2.}

Consistent solutions can be found by taking
\eqn\threefirst{a^0=2,b_1=2,a_1=2.}
This yields
\eqn\threefval{\tau={1\pm i \sqrt{3} \over 2}}
and
\eqn\threefflux{N_{\rm{flux}}=12.}

Alternatively, one can take
\eqn\threesecond{a^0=2,b_1=4, a_1=2.}
In this case,
\eqn\threesval{\tau={1 \pm i\sqrt{7} \over 2}}
and
\eqn\threesflux{N_{\rm{flux}}=28.}
Note that unlike Example 2 above, the two-tori in \threefval\ and
\threesval\ are not square.

Once again, doing general rescalings and $GL(6,{\bf Z}) \times
GL(2,{\bf Z})$ transformations leads to additional solutions in each
of these cases.

As in the previous example, the solutions discussed here have $\ncal
=3$ supersymmetry as well. This follows by the same argument as in the
previous example, after noting that in both the cases \threefirst\ and
\threesecond, $G_{(3)}$ can be expressed as
\eqn\commnttwo{{1\over{(2\pi)^2 \alpha'}}
G_{(3)}=a^0(d\bar{z}^1\wedge dz^2\wedge dz^3).}

\subsec{Toward a general supersymmetric solution}

Solving the supersymmetric equations of motion \susyeqs{}\ without any
simplifying assumptions is a difficult task.  However, a couple of
observations can make the task easier.  First, note that it is
possible to re-write equation \susyeqs{} as
\eqn\CofEq{\Biggl(\cof\biggl(\tau-{1\over A^0}A\biggr)\Biggr)_{ij}
= {1\over {A^0}^2} (\cof A)_{ij} +{1\over A^0}B_{ij}~.}
This determines $\tau^{ij}$ in terms of the flux matrices and the
dilaton $\phi$, since if ${\rm cof}~x = y$, then $x = {\rm cof} y /
\sqrt{\det y}$.  

Next, we note that one can actually eliminate the $\tau^{ij}$ from the
$W=0$ and $\partial_{\tau^{ij}}W = 0$ equations to obtain a quartic
equation for $\phi$.  The quartic is derived in Appendix B, and takes
the form
\eqn\quartic{ (\det A)B_0 - (\det B)A^0 + (\cof A)_{ij}(\cof B)^{ij}
+{\textstyle{1\over4}}(A^0B_0+ A^{ij}B_{ij})^2=0,}
where $A^0=a^0-\phi c^0$, $A^{ij}=a^{ij}-\phi c^{ij}$, and
$B_0,B_{ij}$ are defined similarly.  A quartic equation is soluble, so
one can solve \quartic\ for the allowed values of $\phi$.

This leaves only the equation $\partial_\phi W=0$, which upon
substitution for $\phi$ and $\tau^{ij}$ gives one nonlinear equation
in integers.  The integer equation is a consistency condition that
determines whether the choice of flux can lead to a supersymmetric
solution.  The hard part is solving this equation.  An additional
complication is that for each solution to the integer equation, one
must determine all consistent configurations of exotic orientifold
planes (as described in \joefrey), if one is to find all
supersymmetric solutions.

\newsec{Distinctness of solutions}

Not all solutions with different values of $\phi$ or $\tau^{ij}$ are
physically distinct.  There is an $SL(2,{\bf Z})$ symmetry that
relates equivalent values of the dilaton-axion, and an $SL(6,{\bf Z})$
symmetry that relates equivalent values of the period matrix
$\tau^{ij}$.

\subsec{$SL(2,{\bf Z})$ equivalence}

The type IIB supergravity action \IIBsugra\ is invariant under the
$SL(2,{\bf R})$ symmetry,
\eqn\SLtwotau{\eqalign{
&\pmatrix{F_{(3)}\cr H_{(3)}} \rightarrow \pmatrix{F_{(3)}'\cr
H_{(3)}'} = m \pmatrix{F_{(3)}\cr H_{(3)}},\cr
&\phi \rightarrow \phi' = {a\phi+b\over c\phi+d},\quad
 m=\pmatrix{a & b\cr c & d} \in SL(2,{\bf R}).}}
Under this symmetry, the complex 3-form flux transforms as
\eqn\SLtwoGi{G_{(3)}\rightarrow G'_{(3)} = F_{(3)}'-\phi' H_{(3)}',}
which one can check is equivalent to
\eqn\SLtwoGii{G_{(3)}\rightarrow G'_{(3)} = {G_{(3)}\over c\phi+d}.}
At the quantum level, only an $SL(2,{\bf Z})\subset SL(2,{\bf R})$
survives.  Solutions that differ only by $SL(2,{\bf Z})$
transformations are equivalent.  It is therefore customary to take
$\phi$ to be in the fundamental domain ${\cal F}$, of the upper half
plane modulo $PSL(2,{\bf Z})$:
\eqn\fdomain{{\cal F} = \bigl\{\phi\in{\bf C} \bigm| \Im\phi>0,
-\half\le\Re\phi\le\half,\,|\phi| \ge 1\bigr\}.}
The examples were not chosen in such a way that the solutions would
necessarily give $\phi\in{\cal F}$.  However it is a simple matter to
transform them to the fundamental domain using \SLtwotau, where now
\eqn\SLtwoZ{\pmatrix{a & b\cr c & d}\in SL(2,{\bf Z}).}

\subsec{$SL(6,{\bf Z})$ equivalence}

Following \Schlich, let
\eqn\CthreeBasis{{\cal B}_{{\bf C}^3} =
\bigl(\evec_{(1)},\ \evec_{(2)},\ \evec_{(3)}\bigr)}
denote a basis of ${\bf C}^3$, and consider a $T^6$ in which the
lattice basis is
\eqna\LatticeBasis
$$\eqalignno{{\cal B}_{T^6}
& = \bigl(\evec_{(1)},\ \evec_{(2)},\ \evec_{(3)},\ %
\evec_{(i)}\tau^i{}_1,\ \evec_{(i)}\tau^i{}_2,\ \evec_{(i)}\tau^i{}_3
\bigr)&\LatticeBasis a\cr
&= {\cal B}_{{\bf C}^3}\,\Lambda,\quad\Lambda = \bigl(1,\tau\bigr).
&\LatticeBasis b
}$$
Under a change of lattice basis,
\eqn\Bchange{{\cal B}_{T^6}\rightarrow{\cal B}'_{T^6}={\cal B}_{T^6}M,
\quad M\in SL(6,{\bf Z}),}
so
\eqn\LchangeI{\Lambda\rightarrow\Lambda''=\Lambda M,
\quad M\in SL(6,{\bf Z}).}
The change of lattice basis does not produce $\Lambda''$ in the
standard form $(1,*)$.  However, under a change of ${\bf C}^3$ basis,
\eqn\LchangeII{\Lambda''\rightarrow\Lambda' = N\Lambda'' = N\Lambda M,
\quad N\in GL(3,{\bf C}).}
We can choose $N=N(M,\tau)$, so that $\Lambda'$ is in standard form,
\eqn\LambdaStd{\Lambda'=N\Lambda M=\bigl(1,\tau'\bigr).}
Two period matrices $\tau$ and $\tau'$ related by
\LatticeBasis{b}\ and \LambdaStd, are equivalent.
Also, under an $SL(6,{\bf Z})$ coordinate transformation $M$, the fluxes
$F_{(3)},H_{(3)}$, (when regarded as three-forms) must stay the same
\foot{Under the $SL(6,{\bf Z})$ transformation, \Bchange,
the two coordinate systems are related as: 
\eqn\meanN{M \pmatrix{x_i' \cr y_i'}
=\pmatrix{x_i \cr y_i}.}  
The transformation of $(F_{(3)})_{ijk},(H_{(3)})_{ijk}$ then follow by
requiring that the three forms, $F_{(3)},H_{(3)}$ stay invariant.}.
This means that two solutions with period matrices $\tau$ and $\tau'$
related by \LatticeBasis{b}\ and \LambdaStd, and which are otherwise
identical, are equivalent.

We should make one more comment before turning to an example.  In
Section 7 we discuss solutions which break supersymmetry.  The
analysis above, identifying solutions related by $SL(2,{\bf Z}) \times
SL(6,{\bf Z})$ transformations, applies to these cases as well.

\subsec{Example}

To illustrate the equivalences, consider Example 1 from Section 4.
Suppose instead of choosing the two polynomials \psone, \pstwo, we
made the following choices:
\eqn\newpsone{P_1(\tau)\equiv-(\tau^3+1)=0,}
\eqn\newpstwo{P_2(\tau) \equiv 2\tau^3-3\tau^2+3\tau-1=0.}
These two polynomials have a common factor $P(\tau)=\tau^2-\tau+1$, 
and the corresponding values of integers are 
\eqn\newvalint{((a^0)',a',b',b_0')=(-1,0,0,1) \quad 
((c^0)',c',d',d_0')=(2,1,-1,1),}
where the prime superscripts are being used to distinguish the present
case from Example~1, in Section 4.  Solving $P(\tau)=0$ and choosing
the solution with $\Im(\tau')>0$ gives
\eqn\newvaltau{\tau'=e^{i\pi \over 3}.} 
Also, solving \pthree\ with \newvalint\ gives $\phi'=e^{i2\pi \over
3}$. Finally the total three brane charge in this case is
$N_{\rm{flux}}=3$, as follows from \peight, \newvalint.

This solution is in fact related to the one corresponding to flux,
\pstwelve, by an $SL(6,{\bf Z})$ transformation.

The $SL(6,Z)$ transformation has the form, $S \otimes S \otimes S$
where, each $S \in SL(2,{\bf Z})$, acts on the one of the three
$T^2$'s as:
\eqn\defsl{S: \tau \rightarrow -{1\over \tau}.}
To see this we note first that under \defsl, the modular parameter
$\tau' =e^{i\pi \over 3} \rightarrow e^{2 \pi i \over 3}$, which
agrees with \valtau. Second, one can show that the corresponding
matrix $M$, in \meanN, acting on the coordinates of each $T^2$ has the
form $M=\pmatrix{0 & 1 \cr -1 &0}$. From this it follows that in order
to be related by the same $SL(6,{\bf Z})$ transformation, the flux
integers,
\pstwelve, \newvalint,  must satisfy the conditions:
\eqn\relflux{((a^0)', a',b',b_0') =(-b_0, -b,a,a^0),}
\eqn\refluxb{((c^0)',c',d',d_0')=(-d_0,-d,c,c^0).}
Comparing, \pstwelve, and \newvalint, we see that these conditions are
in fact true.  Finally, the two solutions have the same value for the
dilaton and agree in the value for $N_{\rm{flux}}$. Thus, as per our
general discussion above, they are identical.

\newsec{New Solutions using $GL(2,{\bf Z}) \times GL(6,{\bf Z})$ 
Transformations} 
In various examples of Section 4 we have seen that starting with a
given solution, additional ones can be generated by appropriately
rescaling the fluxes.  Here we discuss this in more generality and
show how additional solutions can be obtained by using $GL(2,{\bf Z})
\times GL(6,{\bf Z})$ transformations.  The resulting solutions are
physically distinct in general, with a different flux contribution to
three brane charge.  Solving the tadpole condition \twostat{}\ without
anti-branes requires that the value of $N_{\rm{flux}}$ for the new
solutions is $\leq 32$, and that the required number of wandering
D3-branes are added in each case.

The general discussion in this section is applied to some examples at
the end. These illustrate that starting with a diagonal period matrix
physically distinct solutions can be obtained with a non-diagonal
period matrix using the $GL({\bf Z})$ transformations.  The examples
also yield solutions where all the three brane charge is cancelled by
fluxes alone, leaving in one instance, four flat directions in K\"ahler
moduli space. These solutions are of the kind mentioned in the
introduction and are good illustrations of the reduced number of
moduli that survive once fluxes are turned on.

\subsec{$GL(2,{\bf Z})$ Transformations}
Consider a solution to the ${\cal N}=1$ susy equations which has flux,
$F_{(3)}$,$H_{(3)}$, and moduli fixed at values $\phi$,$\tau^{ij}$.
Now  transform the fluxes as follows:
\eqn\GLtwotau{\pmatrix{F_{(3)}\cr H_{(3)}} \rightarrow \pmatrix{F_{(3)}'\cr
H_{(3)}'} = m \pmatrix{F_{(3)}\cr H_{(3)}},} where the matrix $m \in
GL(2,{\bf Z})$\foot{By this we mean that $m = \pmatrix{ a & b \cr c &
d}$ where $a,b,c,d \in {\bf Z}$.  In particular $\det(m)$ need not be
$1$.}.

One can show that a solution to the the supersymmetry conditions for
the new fluxes is obtained by taking the moduli to be at the values
\eqn\glvalmod{\phi'={a\phi+b\over c\phi+d}, \quad (\tau^{ij})'=\tau^{ij}.}

To see this note that under the transformation \GLtwotau,
\eqn\newgttr{G_{(3)} \rightarrow G'_{(3)} 
= \det(m) {G_{(3)} \over c \phi + d}.}
The resulting superpotential, \defsuperpota, transforms to
\eqn\gltranssuper{W \rightarrow W'[\phi',\tau]
= \int \Omega \wedge G'_{(3)}
= \det(m) { W[\phi,\tau] \over c \phi + d}.}
where the dependence of the superpotential on the moduli has been
explicitly indicated above.

It now follows that if $W$ satisfied the supersymmetry equations,
\finalcond{}, when the moduli take values $\phi, \tau^{ij}$, then $W'$
will also meet the susy eqautions for the transformed values,
\glvalmod.

Finally, note that under the transformation of the fluxes, \GLtwotau,
the flux contribution to three brane charge becames
\eqn\gltflux{N_{\rm{flux}} \rightarrow N_{\rm{flux}}'= \det(m) N_{\rm{flux}}.}
Starting with a solution, where $N_{\rm{flux}}>0$ we are therefore
restricted to $GL(2,{\bf Z})$ transformations with $\det(m) >0$. Also
as mentioned above, we need to ensure that $N_{\rm{flux}}'\le 32$,
\twostat{}.

\subsec{$GL(6,{\bf Z})$ Transformations}
Our starting point is once again a ${\cal N}=1$ susy preserving
solution with flux, $F_{(3)},H_{(3)}$ and moduli fixed at values
$\phi,\tau^{ij}$.  But this time we consider transforming the flux by
a $GL(6,{\bf Z})$ transformation. The transformation can be described
explicitly as follows.  We fix a basis of one forms $(dx^i,dy^i)$ as
in section 2.4.  The components of $F_{(3)}$ in this basis then
transform as
\eqn\transfls{(F_{(3)})_{abc} \rightarrow (F_{(3)}')_{abc} =
 (F_{(3)})_{def}M^d_aM^e_bM^f_c, }
and similarly for $H_{(3)}$. As a result the components of $G_{(3)}$
in this basis also then transform under $GL(6,{\bf Z})$ as :
\eqn\transgls{(G_{(3)})_{abc} \rightarrow (G_{(3)}')_{abc} =
 (G_{(3)})_{def}M^d_aM^e_bM^f_c.}
In \transfls, \transgls, $M^a_b$ are the elements of a matrix, $M \in
GL(6,{\bf Z})$.

We will see that the new fluxes lead to the moduli being stabilized at
values $\phi',\tau'$ where $\phi'=\phi$ and
\eqn\glstaurel{(1,\tau') = N(1,\tau) M.}
In \glstaurel, $M$ is the same matrix that appears in \transgls, and
$N \in GL(3,C)$ is a matrix that is chosen so that the left hand side
has the form $(1,*)$.  In Appendix C, we show that the superpotential
for the transformed flux, \transgls, is related to the original
superpotential by
\eqn\glzero{W'[\tau',\phi]= \det(N) \det (M) W[\tau,\phi]}
where $\tau',\tau$ are related as in \glstaurel.  It then follows that
if $\tau,\phi$ solve the supersymmetry equations \finalcond{} for the
original fluxes, $\tau',\phi'$ are the solutions for the transformed
fluxes.

Let us also note that under the transformation \transgls, the
contribution to the three brane charge for the new flux is given by
\eqn\glflux{N_{\rm{flux}} \rightarrow N'_{\rm{flux}}
= \det(M) N_{\rm{flux}}.}
Once again we must ensure that the resulting value of three brane
charge meets the consistency checks.

Two more comments are worth making at this stage.  First, suppose the
solution one began with had a diagonal period matrix $\tau$.  Then it
is possible by an appropriate choice of the matrix $M$ to obtain other
solutions where the resulting period matrix $\tau'$, \glstaurel, is
non-diagonal. A specific example will be given in the next section.
Second, in the discussion above we took $M \in GL(6,{\bf Z})$. In
fact, this is not necessary. All that is required is that the
transformed fluxes \transgls, have integer components in the
cohomology basis \basis.\foot{In fact, the coefficients should be even
integers if discrete flux is not being turned on.} For example
choosing $M^a_b=\lambda \delta^a_b$, where $\lambda^3=2$ is perfectly
acceptable.  In this case, we learn from \glstaurel, that $N=\lambda^3
{\bf 1}_{3 \times 3}$, and $\tau'=\tau$. We have already encountered
this case in Section 4.1: doubling the flux rescales the
superpotential and leaves the moduli fixed.

\subsec{An Example}
For an example we start with the a solution discussed in Example 2 of
section 4.3. The fluxes are given by \uneqdo, and the resulting moduli
are stabilized at $\phi=i$ and
\eqn\exvaltau{\tau^{ij}=i \delta^{ij},}
\uneqvalta, \uneqfinaltc, and \uneqeq. 
The solution has $N_{\rm{flux}}=16$.

Now take the matrix $M$, \transgls, to be
\eqn\exvalm{M=\pmatrix{{\bf 1}& 0  \cr
                0& D }.}
Here $M \in GL(6,{\bf Z})$, ${\bf 1}$ is the $3 \times 3$
identity matrix and  $D \in GL(3, {\bf Z})$.    
The resulting values for the fluxes can be worked out using \transgls,
but we will not do so explcitly here. 

The general discussion of the previous section then tells us that the
moduli are stabilized at $\phi'=\phi=i$ and $\tau'$, where $\tau'$ is
given in terms of the original period matrix
\exvaltau\ as discussed in \glstaurel. 
Given $M$ in \exvalm, and $\tau$ in \exvaltau, it is easy to show that
the matrix $N$ in \glstaurel\ is
\eqn\exvaln{N={\bf 1}_{3 \times 3}.}
Therefore,
\eqn\exvaltpr{\tau'= i D. }

The flux contribution to the three brane charge in this case is given by 
\glflux,
\eqn\exflux{N'_{\rm{flux}}=\det(D) N_{\rm{flux}} = 16 \det (D).}
Since $ N'_{\rm{flux}} \le 32$, we learn that $\det(D)=2$ is the only
possibility (cases with $\det(D)=1$ give rise to solutions related to
the original one by $SL(6,{\bf Z})$ transformations, which by the
discussion in section 5.2 are not physically distinct).

As examples for $D$, two possibilities are  
\eqn\exposdo{D=\pmatrix{2 & 0 & 0 \cr 0 & 1 & 0 \cr 0 & 0 & 1 },}
in which case the resulting period matrix is still diagonal \exvaltpr,
but the eigenvalues are unequal.  Or,
\eqn\expodt{D=\pmatrix{1 & -3 & 0 \cr 1 & -1 & 0 \cr 0 & 0 & 1},}
in which case the resulting period matrix is not diagonal.  In the
latter case we see that starting with a diagonal period matrix we have
found an example where $\tau$ is fixed at a non-diagonal value.

It is also useful to briefly revisit the primitivity constraint in the
example \exposdo.  Since the complex structure is diagonal, it is
straightforward to verify that three K\"ahler deformations of the type
\jansatz, survive as flat directions. In addition to these
deformations, the deformation with $w\sim dx^2\wedge dy^3 + dx^3\wedge
dy^2$ is also now of type (1,1).  Thus, altogether there are four
K\"ahler flat directions. This is an example of the kind mentioned in
the introduction.  The fluxes contribute $N_{\rm{flux}} = 32$, so no
extra D3 branes are needed to soak up the orientifold three plane
charge. The dilaton-axion and all complex structure moduli are lifted,
leaving four surviving moduli which are K\"ahler deformations.

\newsec{Solutions with ${\cal N}=2$ Supersymmetry.}
In this section we discuss the conditions which the $G_{(3)}$ flux
must satisfy to preserve $\ncal=2$ supersymmetry. We will illustrate
the discussion with one example at the end of this section. A more
extensive study of $\ncal=2$ preserving vacua is left for the future.

An $\ncal=2$ theory has an $SU(2)_R$ R-symmetry.  $SU(2)_R$ is
embedded in $SO(6)$, the group of rotations along the six dimensional
compactified directions, as follows \foot{This embedding follows by
noting that the spinor $\epsilon$, under which the dilatino and
gravitino variations vanish, must be a doublet of $SU(2)_R$.} :
\eqn\emdedd{SU(2)_R \subset SU(2)_L \times SU(2)_R \subset SO(4) \times U(1)
\subset SO(6).} 
We choose conventions so that the spinor representation, $4$, of
$SO(6)$ transforms as a $(2,1)_{+1} + (1,2)_{-1}$ under $SU(2)_R
\times SU(2)_L \times U(1)$, and the $6$ of $SO(6)$ as $(2,2)_0 +
(1,1)_{+2} + (1,1)_{-2}$.  In the discussion below we will use indices
$a,b$ to denote an element of the $6$ which transforms as $(2,2)_0$
and indices $l,m$ to denote elements transforming as
$(1,1)_2,(1,1)_{-2}$.

Since $SU(2)_R$ is a symmetry of the $\ncal=2$ theory, it must be left
unbroken by the compactification. This means in particular that
$G_{(3)}$ must leave an $SU(2)$ subgroup of $SO(6)$ unbroken.
$G_{(3)}$ transforms as $[6 \times 6 \times 6]_A$ under $SO(6)$. With
respect to $SU(2)_R \times SU(2)_L
\times U(1)$ this decomposes as
\eqn\decom{[6 \times 6 \times 6]_A = (2,2)_0 + (2,2)_0 + (3,0)_2 + (3,0)_{-2}
+(0,3)_{2} + (0,3)_{-2}.}
For $G_{(3)}$ to preserve $SU(2)_R$ it can only have components along
the $(0,3)_{\pm 2}$ representations.  A little thought shows that this
means $G_{(3)}$ has index structure $(G_{(3)})_{abl}$, in the notation
introduced above.

A detailed analysis of the spinor conditions will be presented in
Appendix D.  The conclusion is the following: in order to preserve
$\ncal=2$ supersymmetry $G_{(3)}$ must only take values in the
$(0,3)_{2}$ representation of $SU(2)_R \times SU(2)_L \times U(1)$. In
other words, the $(0,3)_{-2}$ representation, which would have also
preserved $SU(2)_R$, must be absent.

Let us check that this condition on $G_{(3)}$ leads to a solution of
the equations of motion.  The ISD condition \ISDdef, can be written in
the present case as
\eqn\tisd{\epsilon_{abmcdl}G_{(3)}^{cdl}=i(G_{(3)})_{abm},}
which can also be expressed as
\eqn\ttwoisd{\epsilon_{abcd} \epsilon_{ml} G_{(3)}^{cdl}=i(G_{(3)})_{abm}.}
The two $\epsilon$ symbols above refer to the four directions on which
the $SO(4)$ acts and the two directions on which the $U(1)$ acts
respectively.  Since $G_{(3)}$ is a tensor transforming as a $(0,3)$
representation of $SU(2)_R\times SU(2)_L$ it corresponds to the self
dual representation of $SO(4)$ and therefore satisfies the condition
$\epsilon_{abcd} (G_{(3)})^{cdl}=G_{ab}^l$.  Further, one can show, in
our choice of conventions, that a charge $2$ representation of the
$U(1)$ satisfies $\epsilon_{ml} (G_{3)})^{cdl}=i(G_{(3)})^{cd}_m$.
From this, we see that if $G_{(3)}$ is of the $(0,3)_{2}$ kind, it
satisfies the ISD requirement.

It is useful to relate the discussion above to that in section 3.1
where we saw that $G_{(3)}$ must be primitive and $(2,1)$ to preserve
$\ncal=1$ susy.  Requiring $\ncal=2$ supersymmetry must impose extra
conditions on the $G_{(3)}$ flux.  The requirements for $\ncal=1$
supersymmetry mean that under an $SU(3)\subset SO(6)$, $G_{(3)}$
transforms as a ${\bar 6}$. The $SU(2)_L$ discussed above is a
subgroup of this $SU(3)$ (since $\epsilon$ is a singlet under it), so
the ${\bar 6}$ representation of $SU(3)$ transforms under $SU(2)_L$ as
$3+2+1$.  As a result , we learn that $\ncal=1$ susy alone allows
$G_{(3)}$ to take values in the $3+2+1$ representations of $SU(2)_L$.
$\ncal=2$ susy imposes the additional requirement that the doublet and
singlet components are missing and $G_{(3)}$ transforms purely as a
triplet under $SU(2)_L$.  For completeness let us also mention that
for unbroken $\ncal =3$ one must impose yet a further restriction:
$G_{(3)}$ must have only one non-zero component proportional to a
highest weight state of the triplet representation.

These conditions can be visualised as follows.  The weight diagram of
the ${\bar 6}$ representation is a triangle. (See, e.g., (IX.iii) of
\Georgi). Each state in the ${\bar 6}$ representation is denoted by a
point in this diagram. $\ncal =3$ supersymmetry requires $G_{(3)}$ to
be proportional to any one of the three vertices of the triangle,
$\ncal =2$ requires that the components of $G_{(3)}$ all lie along an
edge of the triangle, and finally, $\ncal=1$ supersymmetry allows
components along all six points in the diagram.

In the example below it will be useful to first impose the conditions
for $\ncal=1$ supersymmetry, then check if the extra restrictions for
$\ncal=2$ supersymmetry are met.

\subsec{An ${\cal N}=2$ Example}

As an example choose the fluxes to be:
\eqn\tfluxes{\eqalign{ H_{135}=H_{245}=& F_{136}=F_{246}=
(2 \pi)^2 \alpha' a^0 \cr
F_{135}=F_{245}=& -H_{246}=-H_{136}=(2 \pi)^2 \alpha' a^0,}}
where we are working in the coordinates $x^i,y^i$ introduced in
section 2.4.  Each index above takes six possible values; $i=1,3,5,$
denote components along $x^1,x^2,x^3$ directions, $i=2,4,6,$ along
$y^1,y^2,y^3$.  Also in \tfluxes, $a^0$ is an integer.  In the
cohomology basis, \basis, the fluxes can be expressed as
\eqn\texa{{ 1\over (2 \pi)^2 \alpha'} F_{(3)}= a^0 \alpha_0 + a^0 \beta^0
-a^0 \beta^{33} + a^0 \alpha_{33}}
and
\eqn\texb{ {1\over (2 \pi)^2 \alpha'} H_{(3)}=a^0 \alpha_0 -a^0 \beta^0
-a^0 \beta^{33} -a^0 \alpha_{33}.}

The superpotential is then given by
\eqn\tsuperpot{W=a^0(1-\phi) \det \tau -a^0(1+\phi) (\cof \tau)_{33}
+a^0(1-\phi)\tau^{33} -a^0(1+\phi).}
One can show that the equations for $\ncal=1$ supersymmetry
\susyeqs{}, have the solution
\eqn\ttwosol{\tau^{ij}=i \delta^{ij}, \phi =i.}
The contribution to three brane flux is 
\eqn\tnflux{N_{\rm{flux}}=4(a^0)^2.}
Choosing $a^0=2$ we have $N_{\rm{flux}}=16$ which is within the
acceptable bound, \twostat{}. 

With the choice of complex structure in \ttwosol, $G_{(3)}$ can now be
expressed as

\eqn\expg{{1\over {(2\pi)^2 \alpha'}} 
G_{(3)}={a^0 (1-i) \over 2}(dz^1\wedge d{\bar z}^2 \wedge dz^3
+d{\bar z}^1 \wedge dz^2 \wedge dz^3).}

It is clear that the primitivity condition is satisfied if one chooses
the K\"ahler form to be of the form
\eqn\firstel{J=i\sum_A r_A^2 dz^A \wedge d{\bar z}^A .}
In addition the perturbation
\eqn\secondtel{\delta J=i (dx^1\wedge dy^2 + dx^2\wedge dy^1) \sim
dz^1\wedge d{\bar z}^2 + dz^2 \wedge d{\bar z}^1}
satisfies $\delta J \wedge G = 0$.  The remaining $5$ K\"ahler moduli
are lifted.

So far we have ensured that there is $\ncal=1$ supersymmetry.  We will
now argue that the solution above in fact preserves $\ncal=2$
supersymmetry.

Start by first taking the K\"ahler metric to be $g_{i
\bar\jmath}=\delta_{i \bar\jmath}$. The coordinates $x^i,y^i$ then
define a flat coordinate system. Consider an $SO(4) \times U(1)$
subgroup of $SO(6)$ where the $SO(4)$ acts on the $x^1,x^2,y^1,y^2$,
indices and the $U(1)$ refers to rotations in the $x^3,y^3$, plane. It
is easy to see that for the values \tfluxes, $G_{(3)}$ satisfies the
relation, $\epsilon_{abcd}G_{(3)}^{cdl}=(G_{(3)})_{ab}^l$, and
therefore transforms as a self dual representation of $SO(4)$ (here we
are following the notation of the previous section and the indices
$a,b$ take values $x^1,x^2,y^1,y^2$, while $l,m$ range over
$x^3,y^3$).  Since we have already verified that $G_{(3)}$ satsifies
the $\ncal=1$ conditions, it is ISD, and it follows that it must have
charge $2$ under the $U(1)$.  Putting all this together, in the
example above we find that $G_{(3)}$ transforms as a $(0,3)_2$
representation under $SU(2)_R\times SU(2)_L\times U(1)$.  As per our
discussion above, it therefore meets the requirements for $\ncal=2$
supersymmetry.

Alternatively, working in the complex coordinates $z^i=x^i+i y^i,{\bar
z}^i=x^i-iy^i$, let us define $A_{{\bar k} {\bar l}} =
(G_{(3)})_{ij{\bar k}} \epsilon^{ij}_{\bar l}$.  We see that $A_{{\bar
1} {\bar 1}}$ and $A_{{\bar 2} {\bar 2}}$ have nonzero values in the
above example. Under the $SU(3)$ symmetry, $({\bar z^1}, {\bar z^2},
{\bar z^3}),$ transforms as a ${\bar 3}$ representation. Consider an
$SU(2) \subset SU(3)$ which acts on the ${\bar z}^1,{\bar z}^2,$
coordinates and leaves ${\bar z}^3$ invariant.  $A_{{\bar k}{\bar l}}$
or equivalently $G_{(3)}$ transforms as a triplet of this $SU(2)$.

An additional check, also mentioned in section 4.1, is the following:
in an $\ncal=2$ supersymmetric theory one should be able to define
another inequivalent complex structure which keeps $G_{(3)}$ of kind
$(2,1)$.  In the example above it is easy to see that this corresponds
to choosing holomorphic coordinates $(w^1,w^2,w^3)=({\bar z}^1, {\bar
z^2}, z^3)$.

Finally, some thought shows that under K\"ahler deformations of the
form \firstel, \secondtel, the conditions for $\ncal=2$ supersymmetry
continue to hold.

\newsec{Non-supersymmetric Solutions}

For generic (non-supersymmetric) solutions, we require only that the
scalar potential vanish, or equivalently by \PotAndTension, that
$G_{(3)}$ be ISD.  However, it is computationally simpler to consider
the subclass of solutions in which $G_{(3)}$ is also primitive.  In
this case $G_{(3)}$ can only have pieces of type (2,1) and (0,3).  The
equations that one needs to solve are then

\eqn\nsone{\eqalign{D_{\tau^{ij}} W & = \partial_{\tau^{ij}} W
+ (\partial_{\tau^{ij}} {\cal K}) W = 0,\cr
D_\phi W & = \partial_\phi W + (\partial_\phi {\cal K}) W =
0,}}
along with the primitivity condition,

\eqn\nsprimitive{J\wedge G_{(3)} = 0.}
The first set of equations in \nsone\ imposes the third set of
equations appearing in \fullcond, and forbids type (1,2) pieces of
$G_{(3)}$.  The second equation in \nsone\ is the second equation in
\fullcond, i.e. forbids a (3,0) piece in $G_{(3)}$.  Then, equation
\nsprimitive\ kills the possibility of (2,1) IASD pieces in the
three-form flux ($T^6$, unlike a generic Calabi-Yau, has a
three-dimensional space of IASD non-primitive (2,1) forms).  More
generic non-supersymmetric solutions could be found by relaxing the
requirement that the (1,2) ISD forms be absent from $G_{(3)}$, but we
will not pursue them here.

Fluxes which obey the equations \nsone\ and \nsprimitive\ will break
supersymmetry iff $G_{(3)}$ contains a nontrivial component of type
(0,3).  This is easily interpreted in the low-energy supergravity:
Since we are looking for solutions which are not necessarily
supersymmetric, we no longer need to impose $D_{\rho^\alpha}W \propto
W = 0$ for the K\"ahler moduli, $\rho^\alpha$.  Precisely when
$G_{(3)}$ has a non-vanishing (0,3) piece, $W \neq 0$ and
supersymmetry is broken, but still with vanishing potential (at
leading order in $\alpha'$ and $g_s$).  Examples of such vacua were
discussed in \refs{\GKP,\Beckerst}.  Such vacua will suffer a variety
of instabilities in perturbation theory (as the ``no-scale'' structure
of the potential will be violated by $\alpha'$ and $g_s$ corrections),
which is why we only discuss them briefly here.

The K\"ahler potential for the $\tau^{ij}$ is

\eqn\nstwo{{\cal K} = {\cal K}_{\rm dilaton} + {\cal K}_{\rm cpx}.}

Here,

\eqn\nsthree{{\cal K}_{\rm dilaton} = -\ln \bigl(-i(\phi-\bar\phi)\bigr),}

and

\eqn\three{\eqalign{{\cal K}_{\rm cpx}
&= -\ln\bigl(-i\int_{T^6}\Omega\wedge\bar\Omega\bigr)\cr
&= -\ln\,\det\,\bigl(-i\difftau\bigr)\cr
&=-\ln\bigl(i\epsilon_{ijk}\difftau^{i1}\difftau^{j2}\difftau^{k3}\bigr).}}

Since both $\tau^{ij}$ and $\bar\tau^{ij}$ enter into \nsone, it is in
general difficult to solve the resulting non-holomorphic equations.
However, in an ansatz with enough symmetry, the problem becomes
tractable.

\subsec{A non-supersymmetric example}

Let us make a simple flux ansatz which is a subcase of the ansatz made
in Example 1 of \S4.  We take $a^{ij} = a \delta^{ij}$, $d_{ij} = - a
\delta_{ij}$, and $b_0, c_0$ to be nonzero, with all other fluxes
vanishing.  Then we find that the superpotential takes the form
\eqn\nseone{{1 \over (2\pi)^2 \alpha'} W =
  - c^0 \phi \det \tau -a^{ij} (\cof \tau)_{ij}
  + d_{ij} \phi \tau^{ij} -b_0.}
It is easy to compute the D3-charge carried by the fluxes with this
ansatz,
\eqn\nsetwo{N_{\rm flux}= {1\over(2\pi)^4 (\alpha')^2} \int H_{(3)}
\wedge F_{(3)}=b_0 c^0 - a^{ij}d_{ij}=b_0c^0 + 3a^2.}

From the symmetry of the problem, one can show that $\tau^{ij} =
\tau\delta^{ij}$.  Let us further assume that
\eqn\nsethree{\tau=-\bar\tau,\quad \phi=\tau.}
Then,
\eqn\nsefour{\partial_\tau {\cal K} = -{3\over2\tau},\quad
             \partial_\phi {\cal K} = -{1\over2\tau}, }
so that
\eqn\nsefive{\eqalign{%
D_\tau W
& = \partial_\tau W + (\partial_\tau {\cal K}) W
  = - {3\over2\tau}(c^0\tau^4-b_0)=0,\cr
D_\phi W
& = \partial_\phi W + (\partial_\phi {\cal K}) W
  = - {1\over2\tau}(c^0\tau^4-b_0)=0.
}}

The equations are both satisfied if
\eqn\nsesix{\tau (= \phi) = i \biggl({b_0\over c^0}\biggr)^{1/4},}
therefore our assumption was consistent.  Finally, since the flux
ansatz is a special case of \S4\ Example 1, we can solve \nsprimitive\
by taking $J$ to be in the same space that led to $G_{(3)}$ primitive
in Section 4.1.  We can also check that the conditions for
supersymmetry here are the same as those found earlier.  The solution
will be supersymmetric if $W=0$.  In the present example,
\eqn\nseseven{W = - 6a\tau^2-2b_0 = 2b_0
\biggl(\sqrt{{9a^2\over b_0c^0}}-1\biggr).}
So, the solutions are non-supersymmetric as long as $9a^2 \neq b_0c^0$. 
In fact, it turns out there are no solutions which have even 
fluxes, $9a^2 = b_0 c^0$ and
$N_{\rm{flux}} \leq 32$ in any case. 

\newsec{Brane Dynamics}

In many of the examples of ${\cal N}=1$ vacua with flux, one finds
that the number of space-filling D3 branes needed to satisfy the
tadpole cancellation requirement \twostat{}\ is
\eqn\nbranes{N_{\rm D3} = 16-\ha N_{{\rm O3}'} - 
{1\over 2(2\pi)^4 (\alpha')^2}\int
H_{(3)} \wedge F_{(3)} ~>~ 0}
($N_{D3} \geq 0$ is needed for supersymmetry).  Therefore, in
addition to the background 3-form flux, one must introduce
space-filling D3 branes.

Following the work of Myers \Myers, it has been recognized that
background p-form fields can have interesting effects on brane
dynamics.  It follows from \Myers\ that the worldvolume potential
(working at vanishing RR axion $C_0$) is given by
\eqn\vh{{\cal V}_{\rm open} \sim {1\over g_s} H_{ijk} Tr(X^i X^j X^k) -  
(*_6 F_{(3)})_{ijk} Tr(X^i X^j X^k) + \cdots}
where $\cdots$ includes the usual ${\cal N}=4$ field theory potential.
When $G_{(3)}$ is ISD,
\eqn\truefact{*_6 F_{(3)} ~=~{1\over g_s} H_{(3)}}
and the first two terms in \vh\ exactly cancel.  

This is in keeping with the fact that the ISD fluxes mock up D3 brane
charge and tension, and satisfy a ``no force'' condition with the D3
branes \GKP.  Therefore, at least at large radius (where supergravity
intuition applies), the D3 point sources are free to live at arbitrary
positions on the $T^6$.  When $k \leq N_{D3}$ branes meet at a generic
point, the low-energy physics is that of $SU(k)$ ${\cal N}=4$ SYM
theory, while $k$ branes meeting at an O3 plane will give rise to an
$SO(2N)$ theory, as usual.  It would be interesting to determine the
leading nontrivial effects of the fluxes on the D-branes, and to find
more elaborate types of models where phenomena reminiscent of those
observed in \PS\ can occur.  Inclusion of anti-branes in the flux
background might also lead to interesting phenomena, as in
\KPV. 

It follows from this discussion that inclusion of $N_{D3}$ branes in
one of our models adds $3 N_{D3}$ complex moduli to the low energy
theory.  From this perspective, the models with $N_{\rm{flux}} \simeq
32$ and $N_{D3} \simeq 0$ are the most satisfying.

\newsec{Discussion}

IIB compactifications on Calabi-Yau spaces with both RR and NS 3-form
fluxes turned on provide a rich class of vacua which are amenable to
detailed study.  It should be clear that the techniques used here to
compute $W$ and study vacua of the $T^6/Z_2$ orientifold would
generalize to many other examples.  The main novelty of these examples
is that they provide a setting where the stabilization of Calabi-Yau
moduli becomes a concrete and tractable problem.  These models are
also of interest because they give rise to warped compactifications of
string theory, and in some cases the low-energy physics has a
holographic interpretation via variants of the AdS/CFT duality
\refs{\Herman,\GKP}.

Several natural questions about the $T^6/Z_2$ models studied here
would be suitable for further study.  A complete classification of
supersymmetric vacua may be possible (although, especially in cases
where the additional complications of discrete RR and NS flux arise
\joefrey, it could be very difficult to achieve).  It is also
interesting to ask whether there are any cases where, with a fixed
topological class for the fluxes, one finds multiple vacua.  Finally,
various dual descriptions of these models should exist, and fleshing
out these dualities (and in particular, understanding any analogues of
mirror symmetry for vacua with nonzero $H$-flux) seems worthwhile.

\medskip
\centerline{\bf{Acknowledgements}}

We would like to thank S.~Das, K.~Dasgupta, S.~Gukov, R.~Kallosh,
N.~Kaloper, J.~Polchinski, S.~Sethi, E.~Silverstein, P.~Tripathi and
H.~Verlinde for useful discussions about related subjects.  S.T. is
particularly grateful to A.~Dabholkar for several helpful comments and
insights and to T.~R~Ramadas for bringing \Hitchin\ to his attention
and for related discussions.  S.K. enjoyed the hospitality of the
Weizmann Institute of Science and the Amsterdam Summer Workshop on
String Theory during the course of this work.  This research was
supported in part by the Department of Energy under contract
DE-AC03-76SF00515.  The work of S.K. is also supported by a Packard
Fellowship, a Sloan Fellowship, and National Science Foundation grant
PHY-0097915.  S.T. acknowledges support from the IMSC String workshop,
2001, where some of this work was done.

\medskip

\appendix{A}{Flux quantization}

We follow the conventions of \GKP\ and \JoeTwo.  A Dp-brane couples to
the $(p+2)$-form RR field strength via the action
\eqn\appAone{-{1\over2\kappa_{10}^2}{1\over2(p+2)!}\int_{{\cal M}_6} 
d^{10}x\sqrt{-g}\ F_{(p+2)}{}^2 + \mu_p\int C_{p+1}.}
The usual quantization condition that follows from this action is
\eqn\appAtwo{\int_\gamma F_{p+2} = \bigl(2\kappa_{10}{}^2\mu_{6-p}
\bigr)n_\gamma,
\quad n_\gamma\in {\bf Z},
\quad\mu_p={1\over(2\pi)^p}\alpha'^{-{p+1\over2}},}
for an arbitrary 3-cycle $\gamma\in H_3({\cal M}_6,{\bf Z}).$ Here
$\mu_p$ is the electric charge of a D$p$-brane and $\mu_{6-p}$ is the
charge of the dual magnetic D$(6-p)$-brane.  The product of these two
charges is related to the factor $1/2\kappa_{10}{}^2 =
(2\pi)^7\alpha'^4$ that multiplies the action, via the Dirac
quantization condition
\eqn\appAthree{\mu_p \mu_{6-p} = {2\pi\over2\kappa_{10}{}^2}.}
From \appAtwo\ and \appAthree,
\eqn\appAfour{\mu_p \int F_{p+2} = 2\pi n,\quad n\in {\bf Z},}
which, in the case $p=1$, becomes
\eqn\appAfive{{1\over2\pi\alpha'}\int F_3 = 2\pi n,\quad n\in {\bf Z}.}
Similarly, we know that the electric NS charge of a fundamental string
is $\mu_{\rm F1}=1/2\pi\alpha'$.  So, using $\mu_{\rm F1}\mu_{\rm NS5}
= 2\pi/2\kappa_{10}{}^2$ together with the analog of the first
equation in \appAtwo,
\eqn\appAsix{{1\over2\pi\alpha'}\int H_3 = 2\pi n,\quad n\in {\bf Z}.}
This equation can also be obtained from \appAfive\ by S-duality.

For compactification on $T^6/Z_2$, it can be shown that the
quantization condition is exactly \appAtwo, with ${\cal
M}_6=T^6$\joefrey\foot{We are indebted to A.~Frey and J.~Polchinski
for providing us with a preliminary draft of their preprint\joefrey.
The remainder of this section summarizes an analogous section in their
preprint.}.  The 3-cycles on $T^6/Z_2$ include both the 3-cycles on
$T^6$ and also new cycles, such as
\eqn\appAseven{\gamma_0\colon 0\le x^1,x^2\le1,\quad 0\le x^3\le\ha,
\quad y^i=0,}
which are ``half-cycles'' on $T^6$.  Naively, this would seem to lead
to a problem with the quantization condition \appAtwo.  Define
$\gamma_1$ by
\eqn\appAeight{\gamma_1\colon 0\le x^1,x^2,x^3\le1,\quad y^i=0.}
Then, one has $n_{\gamma_0}=\half n_{\gamma_1}$, so that
$n_{\gamma_0}\not\in {\bf Z}$ when $n_{\gamma_1}$ is odd.  However, as
discussed in \joefrey, the quantization condition is still satisfied
in this case, if a half unit of discrete RR flux is turned on at an
odd number of the O3-planes that intersect $\gamma_{0,1}$.  Similarly,
when $m_{\gamma_1}$ is odd, a half unit of NS flux must be turned on
at an odd number of the O3-plane that intersects $\gamma_{0,1}$.  When
$n_{\gamma_1}$ ($m_{\gamma_1}$) is even, it is also permissible to
turn on RR (NS) flux at some of the O3-planes that intersect
$\gamma_{0,1}$, but we require that the total number of such O3-planes
be even.  Because the construction of vacua with these exotic O3
planes is somewhat involved except in the simplest examples, we have
focused in this paper on cases where all of the fluxes in the covering
space are even integers, and the naive problem does not arise.

\appendix{B}{Derivation of equation \quartic}
{ 
\def\tilde{\widetilde}

Write $\tau^{ij}=T^{ij}+A^{ij}/A^0=T^{ij}+\tilde A^{ij}$, where a
tilde denotes division by $A^0$. Here, $A^{ij}=a^{ij}-\phi c^{ij}$,
and $B_{ij}$, $A^0$ and $B_0$ are defined similarly.  Then, equation
\susyeqs{c}\ becomes
\eqn\QuarticOne{\tilde W = \det\tau - \tilde A^{ij}(\cof\tau)_{ij}
- \tilde B_{ij}\tau^{ij} - \tilde B_0,}
which, after some algebra can be shown to have the $T^{ij}$ expansion
$$\tilde W = \det T - \bigl( (\cof\tilde A)_{ij} + \tilde B_{ij}
\bigr) T^{ij} - \bigl( \tilde A^{ij}\tilde B_{ij} + \tilde B_0 +
2\det\tilde A \bigr).$$
The analog of equation \susyeqs{c} has already been
obtained in equation \CofEq,
\eqn\QuarticTwo{(\cof T)_{ij} = (\cof\tilde A)_{ij} + \tilde B_{ij}.}
By virtue of this equation, the previous result becomes
$$\tilde W = - 2\det T + (\tilde A^{ij}\tilde B_{ij}
+ \tilde B_0 + 2\det\tilde A).$$
When $W=0$,
\eqn\QuarticThree{\det T = -\ha(\tilde A^{ij}\tilde B_{ij} + \tilde B_0
+2\det\tilde A).}
Since we have independent expressions \QuarticTwo\ and \QuarticThree\
for $\cof T$ and $\det T$, respectively, the equality
\eqn\QuarticFour{\det\cof T = (\det T)^2}
gives a quartic equation for $\phi$.  Explicitly, we have
\eqn\QuarticFive{\eqalign{
\det\cof T &= \det\cof\tilde A + (\cof\cof\tilde A)^{ij}\tilde B_{ij}
+(\cof\tilde A)_{ij} (\cof\tilde B)^{ij} +\det\tilde B\cr
&= (\det\tilde A)^2 +(\det\tilde A)(\tilde A^{ij}\tilde B_{ij})
+ (\cof\tilde A)_{ij}(\cof\tilde B)^{ij} + \det\tilde B,}}
and
\eqn\QuarticSix{(\det T)^2 = (\det\tilde A)^2
+ (\det\tilde A)(\tilde A^{ij}\tilde B_{ij} + \tilde B_0)
+{\textstyle{1\over4}}(A^0B_0+ A^{ij}B_{ij})^2,}
so, equating the two and multiplying by $(A^0)^4$,
\eqn\QuarticSeven{(\det A)B_0 - (\det B)A^0 + (\cof A)_{ij}(\cof B)^{ij}
+{\textstyle{1\over4}}(A^0B_0+ A^{ij}B_{ij})^2=0,}
which is equation \quartic.

} 

\appendix{C}{Derivation of Equation \glzero}
To establish that \glzero, is correct, notice first that the
transformed superpotential for the new fluxes is given by
\eqn\glssuper{W'[\tau]=\int G'_{(3)} \wedge \Omega[\tau],}
where we have explicitly indicated that the dependence on the
complex structure moduli arises from $\Omega$ on the right hand side.
Using, \transgls, \glssuper, can also be expressed as
\eqn\secglssuper{ W'[\tau]=(G_{(3)})_{rst}
\Omega[\tau]_{def}M^r_aM^s_bM^t_c
\epsilon^{abcdef}.}
Now,
\eqn\defholoth{\Omega[\tau]=dz^1\wedge dz^2\wedge dz^3,}
where,
\eqn\glone{\pmatrix{dz^1\cr dz^2\cr dz^3}
=(1, \tau) \cdot \pmatrix{dx^i \cr dy^i}.}  
Under a change of complex structure, $\tau \rightarrow \tau'$ (where
$\tau'$ is given by \glstaurel)
\eqn\glallone{\pmatrix{dz^1\cr dz^2\cr dz^3} \rightarrow
\pmatrix{(dz^1)' \cr (dz^2)'
\cr (dz^3)'} =  N (1,\tau) M  \cdot \pmatrix{dx^i \cr dy^i}.}
As a result one finds that\foot{This follows, for example, by noting
from \glallone\ that, up to an overall normalization of $\det(N)$,
$\Omega[\tau']$ in the basis $\pmatrix{(dx^i)' \cr (dy^i)'}= M
\pmatrix{dx^i \cr dy^i}$ has the same components as $\Omega[\tau]$ in
the basis $\pmatrix{dx^i \cr dy^i}$.}
\eqn\changehol{\Omega[\tau']_{(def)}=
\det(N) \Omega[\tau]_{uvw}M^u_dM^e_bM^w_f.}
Substituting in \secglssuper, then leads to
\eqn\glthree{W'[\tau']=\det(N) \det(M) W[\tau].}

\appendix{D} {The spinor  conditions for ${\cal N}=2$ Supersymmetry}
Throughout this appendix the components for all tensors will be
evaluated in a vielbein frame. We will also use the notation
introduced in section 7.  The $SO(6)$ group of rotations in the $6$
compactified directions has an $SO(4) \times U(1)$ subgroup. In our
notation, indices $a,b $ which take four values refer to directions
which transform under the $SO(4)$ and indices $l,m$ which take two
values refer to directions which are acted on by the $U(1)$ subgroup.
The metric in the vielbein frame has components $g_{ab}=\delta_{ab},
g_{lm}=\delta_{lm}, g_{al}=0$. Also the $\gamma$ matrices satisfy the
relation
\eqn\tspgamma{\{\gamma^l,\gamma^a\}=0.}

Using the fact that an $SU(2)_R$ symmetry group must be left unbroken
we argued in section 7 that the flux must have the index structure,
$(G_{(3)})_{(abl)}$, and further that $G_{(3)}$ must transform as
$(0,3)_{\pm 2}$ under the $SU(2)_R\times SU(2)_L\times U(1) \subset
SO(6)$ group.  Here we will show that the spinor conditions imply that
the $(0,3)_{-2}$ terms must be absent and $G_{(3)}$ must only
transform as a $(0,3)_2$ representation under this group.

The spinor conditions are given in \GPOne\ and \spsix,
\eqn\tspone{G_{(3)} \epsilon = G_{(3)} \epsilon^{*} = G_{(3)} \gamma^{l}
\epsilon^{*}=
G_{(3)} \gamma^a \epsilon^{*} =0.}

In our choice of conventions, the spinor $4$ representation of $SO(6)$
transforms as $(2,1)_1 + (1,2)_{-1}$ under $SU(2)_R \times SU(2)_L
\times U(1)$.  In the $\ncal=2$ supersymmetry case, $\epsilon$ is a
doublet of $SU(2)_R$ and therefore transforms as a $(2,1)_1$
representation of $SU(2)_R\times SU(2)_L\times U(1)$.  We are now
ready to ask what conditions \tspone\ imposes on the flux $G_{(3)}$.

We noted above that the flux has index structure $G_{abl}$.  Using
\tspgamma, the first condition in \tspone\ can be explicitly written
as
\eqn\tsptwo{(G_{(3)})_{lab}\gamma^l [\gamma^a, \gamma^b ] \epsilon=0.}
If $G_{(3)}$ transforms as $(0,3)_{\pm 2}$ under $SU(2)_R \times
SU(2)_L \times U(1)$ it is easy to see that $(G_{(3)})_{lab}[\gamma^a,
\gamma^b ]$ is a generator of $SU(2)_L$ and therefore must annihilate
$\epsilon$, which is a singlet of $SU(2)_L$.  So \tsptwo\ is met.

Similarly, since $\epsilon^{*}$ is also a singlet under $SU(2)_L$, it
is also true that $G_{(3)} \epsilon^{*}=0$.  The third condition in
\tspone, can be written as
\eqn\tthird{{1 \over 2} (G_{(3)})_{mab}\gamma^m\gamma^l [\gamma^a,\gamma^b]
\epsilon^{*}=0.}
Once again the same argument leading to the first two conditions being
met ensures that \tthird\ is also satisfied.

Finally we come to the last condition in \tspone.  This can be
expressed as
\eqn\tlast{(G_{(3)})_{bcl}\gamma^b\gamma^c\gamma^a\gamma^l \epsilon^{*}=0.}
One can show that \tlast\ is not met if $G_{(3)}$ has a $(0,3)_{-2}$
component. If this component is absent though, and $G_{(3)}$ is
entirely of the $(0,3)_{2}$ kind, one can show that
\eqn\tspfive{(G_{(3)})_{bcl}\gamma^l\epsilon^{*}=0.}
Condition \tlast\ then follows. 

To show that \tspfive\ is satisfied when $G_{(3)}$ transforms as a
$(0,3)_{2}$ state we first note, as was pointed out above, that
$\epsilon$ has charge $+1$ with respect to the $U(1)$.  So
$\epsilon^{*}$ has charge $-1$.  As a result, if $G_{(3)}$ is of
$(0,3)_{2}$ kind, $(G_{(3)})_{abl}\gamma^l\epsilon^{*}$, has charge
$-3$ under the $U(1)$.  Also note that the state $(G_{(3)})_{abl}
\gamma^l\epsilon^{*}$ transforms as a $4$ spinor under the $SO(6)$
symmetry. But the $4$ representation does not have any state with $-3$
charge under the $U(1)$ symmetry. Thus the left hand side of \tspfive\
must vanish.

In summary, the spinor conditions show that $G_{(3)}$ must transform
under $SU(2)_R \times SU(2)_L \times U(1)$ as a $(0,3)_{2}$
representation, in order to preserve $\ncal=2$ supersymmetry.

\listrefs
\bye